\documentclass[preprint,preprintnumbers,amsmath,amssymb,titlepage]{revtex4-2}
\usepackage{graphicx}
\usepackage{comment}
\usepackage[normalem]{ulem}
\usepackage{bm}
\usepackage{color}
\usepackage[symbol,bottom]{footmisc}

\usepackage{dcolumn}
\usepackage[left]{lineno}
\newcommand{\FST}{FeSe$_x$Te$_{1-x}$}

\newcommand{\Hc}{$H_{\rm c2}$}
\newcommand{\Hco}{$H_{\rm c2}^{\rm orb.}$}
\newcommand{\HcP}{$H_{\rm c2}^{\rm P}$}
\newcommand{\tc}{$T_{\rm c}$}

\begin{document}

\title{
Non-Fermi liquid transport and strong mass enhancement near the nematic quantum critical point in FeSe$_x$Te$_{1-x}$ thin films
}

\author{Yuki Sato$^{1}$}
\email{yuki.sato.yj@riken.jp}
\author{Soma Nagahama$^{2}$}
\author{Ilya Belopolski$^{1}$}
\author{Ryutaro Yoshimi$^{1,3}$}
\author{Minoru Kawamura$^{1}$}
\author{Atsushi Tsukazaki$^{2,4}$}
\author{Akiyoshi Yamada$^{1,5}$}
\author{Masashi Tokunaga$^{1,5}$}
\author{Naoya Kanazawa$^{6}$}
\author{Kei S. Takahashi$^{1}$}
\author{Yoshichika $\bar{\rm O}$nuki$^{1}$}
\author{Masashi Kawasaki$^{1, 2}$}
\author{Yoshinori Tokura$^{1, 2, 7}$}

\affiliation{$^{1}$RIKEN Center for Emergent Matter Science (CEMS), Wako 351-0198, Japan} 
\affiliation{$^{2}$Department of Applied Physics and Quantum-phase Electronics Center (QPEC), University of Tokyo, Tokyo 113-8656, Japan}
\affiliation{$^{3}$Department of Advanced Materials Science, University of Tokyo, Kashiwa 277-8561, Japan}
\affiliation{$^{4}$Institute for Materials Research (IMR), Tohoku University, Sendai 980-8577, Japan}
\affiliation{$^{5}$Institute for Solid State Physics, University of Tokyo, Kashiwa 277-8581, Japan}
\affiliation{$^{6}$Institute of Industrial Science, University of Tokyo, Tokyo 153-8505, Japan}
\affiliation{$^{7}$Tokyo College, University of Tokyo, Tokyo 113-8656, Japan}

\date{\today}
\maketitle

\begin{center}
    \textbf{Abstract}
\end{center}
Unconventional superconductivity is often accompanied by non-Fermi liquid 
(NFL) behavior, which emerges near a quantum critical point (QCP) \textemdash a point where an electronic ordered phase is terminated at absolute zero under non-thermal parameters. While nematic orders, characterized by broken rotational symmetry, are sometimes found in unconventional superconductors, the role of nematic fluctuations in driving NFL transport behavior remains unclear. Here, we investigated electrical and thermoelectric transport properties in FeSe$_x$Te$_{1-x}$ thin films and observed hallmark NFL behavior: temperature-linear resistivity and logarithmic divergence of thermoelectricity at low temperatures. Notably, the thermoelectricity peaks sharply at the nematic QCP ($x$ = 0.45), highlighting the dominant role of nematic fluctuations in the NFL transport. Furthermore, we found that the pair-breaking mechanisms in the superconducting phase crosses over from orbital- to Pauli-limited effects, indicating the mass enhancement near the nematic critical regime. These findings reveal the profound impact of nematic fluctuations on both normal-state transport and superconducting properties.

\newpage

One of the central challenges in researching unconventional superconductivity is understanding the role of quantum fluctuations in non-Fermi liquid (NFL) transport near a quantum critical point (QCP), where a second-order phase transition takes place at absolute zero \cite{Lohneysen2007Fermi, shibauchi2014quantum, paschen2021quantum, phillips2022stranger}. In some strongly correlated superconductors, such as cuprate \cite{sato2017thermodynamic}, iron-based superconductors \cite{fernandes2022iron}, layered kagome superconductor \cite{nie2022charge}, and magic-angle twisted bi-layer graphene \cite{cao2021nematicity}, an electronic nematic state has been observed, where the electronic system breaks the rotational symmetry while preserving the underlying lattice translational symmetry. Investigating transport properties near a QCP associated with such nematic orders is of great importance, as it may be linked to the symmetry of the superconducting order parameter and the microscopic pairing mechanisms. In the vicinity of a QCP, strong renormalization of quasiparticle mass $m^{*}$ has been reported \cite{shibauchi2014quantum, Ramshaw2015Quasiparticle}, attributed to enhanced electron correlations by quantum fluctuations. However, because nematic orders often coexist or compete with other electronic orderings such as magnetism and charge-density waves, accessing a pure nematic QCP, isolated from the other order, poses a significant challenge. Consequently, it remains unclear whether nematic fluctuations alone are responsible for the observed NFL transport behavior and mass enhancement in strongly correlated superconductors.

Iron-chalcogenide superconductors have attracted considerable attention due to their unique electronic properties owing to the strongest electronic correlations among the iron-based superconductor family \cite{shibauchi2020exotic, yin2011kinetic, yi2017role}. FeSe stands out in this family for exhibiting an electronic nematic phase in the absence of magnetism. Isovalent substitution of Se with Te in a form of FeSe$_x$Te$_{1-x}$ (FST) significantly alters its electronic structure in several ways. First, it monotonically suppresses the nematic order, with the transition temperature $T_{\rm nem}$ reaching absolute zero at around $x$ = 0.45, leading to a nematic QCP \cite{mukasa2021high} (Fig. \ref{fig:RT}(a)). Correspondingly, nematic susceptibility shows a pronounced divergence at around $x$ = 0.5 \cite{ishida2022pure, jiang2023nematic}. Second, Te-substitution enhances the interlayer Te 5$p_z$-5$p_z$ hybridization owing to a large ionic radius of Te compared to Se, and reduces the intralayer hopping within the iron-chalcogenide plane. This leads to the enhancement of electronic correlations within the layer \cite{yin2011kinetic, yi2017role}. In the end compound FeTe, a double-stripe antiferromagnetic (AFM) state emerges with an ordering vector $Q$ = ($\pi$, $\pi$) below N$\acute{\rm e}$el temperature $T_{\rm N} \sim$ 75 K \cite{bao2009tunable}. This AFM phase is steeply suppressed as Te content decreases, resulting in another potential QCP associated with the disappearance of AFM order at around $x$ = 0.08 \cite{otsuka2019incoherent} (Fig. \ref{fig:RT}(a)). Notably, nuclear magnetic resonance experiments at $x$ = 0.42 reveal a nearly temperature ($T$)-independent spin-lattice relaxation rate, suggesting the absence of significant AFM fluctuations around the nematic QCP \cite{arvcon2010coexistence}. Accordingly, the nematic QCP is isolated from the AFM QCP, making FST an ideal materials platform for studying the intrinsic physics associated with this pure nematic QCP. Recently, high magnetic field transport measurements in an analogous compound Fe(Se,S) revealed $T$-linear resistivity extending over a decade in temperature near a nematic QCP, suggesting a connection between nematic fluctuations and NFL transport \cite{licciardello2019electrical}. However, it remains unclear whether nematic fluctuations can induce mass enhancement at the pure nematic QCP \cite{lederer2017superconductivity, coldea2019evolution, mukasa2023enhanced}. 

One of the challenges in studying the systematic transport properties of FST is synthesis of high-quality homogeneous crystals across the entire $x$ range. Solid-state reaction methods results in phase-separation within a certain composition range (0.6 $\leq x \leq$ 0.9), while the chemical vapor transport technique can produce high-quality single crystals, but only for Se-rich compositions (0.5 $\leq x \leq$ 1) \cite{mukasa2021high}. Additionally, one has to carefully treat them with annealing procedure, which is believed to remove excess interstitial Fe atoms and enhances superconductivity \cite{sun2016influence}. In contrast, the molecular beam epitaxy (MBE) technique enables the fabrication of FST thin films across the entire $x$ range without phase separation \cite{Sato2024Molecular}. MBE also offers a controllable in-situ Te-annealing and capping process, which is useful in preserving modest superconductivity \cite{Sato2024Molecular}. Practically, MBE allows relatively short fabrication time (several hours per composition) compared to that typically required for bulk crystal growth, enabling the production and characterization of dozens of films in this study. These advantages make MBE a powerful tool for studying the intrinsic electronic properties across the phase diagram of FST.

In this paper, we present an experimental study of electrical resistivity $\rho$, Seebeck coefficient $S$, and upper critical field \Hc\ for MBE-grown FST thin films spanning a wide range of $x$ across the nematic QCP. We observed hallmark NFL transport behavior in both resistivity and Seebeck coefficient: $T$-linear resistivity and a logarithmic divergence of Seebeck coefficient, $S/T \propto \ln{T}$, near the nematic QCP at $x$ = 0.45. These results support the scenario in which nematic fluctuations drive NFL transport behavior in strongly correlated metals. Additionally, we observed a crossover in the dominant pair-breaking mechanisms from orbital to Pauli effect around the nematic QCP, indicative of a mass enhancement due to nematic quantum criticality. Our results thus point to the strong connection between nematic fluctuations and NFL transport properties and their impact on superconducting pair-breaking mechanisms.

We investigate the electrical resistivity $\rho$ of FST films with a wide composition range of $0 \leq x \leq 0.72$, as a function of $T$ (Fig. \ref{fig:RT}(b)). As $T$ decreases, $\rho$ increases at high temperatures, peaks at temperatures $T_{\rho}^{*}$ (as indicated by the arrows in Fig. \ref{fig:RT}(b)), turns to decrease with a strong suppression of $\rho$ at low temperatures, and finally drops to zero at the superconducting transition temperature \tc, except for the non-superconducting FeTe ($x$ = 0) (Fig. \ref{fig:RT}(c)). The overall behavior is similar to that of Te-annealed bulk FST crystals with minimal excess iron \cite{sun2016influence}. This similarity suggests that the content of excess irons is successfully reduced in the MBE-grown FST thin films processed with Te-annealing.

Remarkably, the $\rho$-$T$ curves for the FST thin films can be classified into two categories based on their curvatures at low temperatures: one with $T^2$ dependence and the other with $T$-linear dependence. In the former category, including $x$ = 0, 0.49, and 0.72, $\rho$ continues to decrease with a downward concavity as $T$ decreases, and it appears to saturate at low temperatures. This behavior is characteristics of Fermi liquid (FL) transport, and the $\rho$-$T$ curves in this category can be fitted with $\rho = \rho_0 + A_{2}T^2$ at sufficiently low temperatures, as shown by the black dashed lines in Fig. \ref{fig:RT}(d). In the other category, including $x$ = 0.06, 0.31, and 0.45, $\rho$ decreases steeply with an upward concavity without saturation down to \tc. The upward concavity corresponds to a low power $n \leq 1$ in a empirical model $\rho = \rho_0 + AT^n$, suggesting the absence of FL transport down to \tc. Notably, as shown in the red dashed lines in Fig. \ref{fig:RT}(d), the $\rho$-$T$ curves for this category exhibits $T$-linear resistivity ($\rho = \rho_0 + A_{1}T$) at sufficiently low temperatures but above \tc, reminiscent of quantum critical metals near AFM QCPs, such as heavy fermions \cite{Lohneysen2007Fermi} and cuprates \cite{legros2019universal}. We note that the value of $\rho$ for all the samples investigated is comparable to, or even exceeds the Mott-Ioffe-Regel (MIR) limit of $\rho_{\rm MIR} \sim$ 0.8 m$\Omega$cm for FST, where the mean free path $\ell$ becomes extremely short and reaches the lattice constant (see Supplemental Materials \cite{SM} for more details). This indicates that FST films lie within the \textit{bad metal} regime \cite{phillips2022stranger}, suggesting that the large value of $\rho$ is not dominated by extrinsic disorder but is likely an intrinsic feature driven by strong quantum fluctuations.

Thermoelectricity is another highly sensitive probe for detecting quantum criticality in the normal state of unconventional superconductors. It is widely acknowledged that the Seebeck coefficient divided by temperature, $S/T$, remains constant within the framework of FL theory. In contrast, in the NFL regime, $S/T$ can exhibit a logarithmic divergence with approaching an AFM QCP, as given by $S/T \sim A_S\ln{T}$ \cite{Paul2001Thermoelectric}, where $A_S$ serves as a measure of quantum fluctuations (see Supplemental Materials \cite{SM} for more details). 

Figure \ref{fig:ST}(a) shows the $T$-dependence of $S$ for the series of FST thin films with various $x$. The non-$T$-linear behavior of $S$ across the entire temperature range cannot be explained solely within the framework of the FL theory. To assess the NFL transport properties in the thermoelectricity, we plot $S/T$ as a function of the logarithmic scale of $T$ (Figs. \ref{fig:ST}(b)-(g)). At high temperatures above approximately 100 K, $S$ shows an almost $T$-independent positive value, except for the negative $S$ in $x$ = 0. This behavior is consistent with Heikes formalism, where strong on-site Coulomb interactions lead to $T$-independent $S$ \cite{pallecchi2016thermoelectric}. Upon decreasing $T$, $S/T$ begins to decrease at $T_{S}^{*}$ (Fig. \ref{fig:ST}(h)), exhibits a sign change from positive to negative, and shows a logarithmic increase over a temperature range typically spanning 10-50 K, as indicated by the solid lines in Figs. \ref{fig:ST}(b)-(g). Further decrease in $T$ results in an abrupt reduction in the amplitude of $S/T$ at \tc\ due to the superconducting transition, except the non-superconducting $x$ = 0. Additionally, $S/T$ saturates at low temperatures for $x$ = 0 and 0.72 (the dotted lines in Figs. \ref{fig:ST}(b) and \ref{fig:ST}(g)), consistent with the formation of FL statea, as evidenced by the $T^2$ dependence of $\rho$.

The observed logarithmic divergence of $S/T$ in FST thin films \cite{Xu2023In} is reminiscent of similar behavior reported in various quantum critical metals near AFM QCPs, including heavy fermions \cite{Paul2001Thermoelectric}, cuprates \cite{daou2009thermopower, Mandal2019anomalous}, and iron arsenides \cite{pallecchi2016thermoelectric, arsenijevic2013signatures}. At AFM QCPs, this logarithmic divergence arises from enhanced scattering processes mediated by spin fluctuations, which leads to $T$-dependent mass renormalization, $m^{*}/m \sim \ln(1/T)$ \cite{Paul2001Thermoelectric}. Such mass enhancement is often evidenced by a strong enhancement of the electronic specific heat near magnetic QCPs \cite{shibauchi2014quantum}. Similarly, the logarithmic divergence of $S/T$ near the nematic QCP in FST suggests a strong connection between nematic fluctuations and the observed NFL transport behavior, accompanied by strong mass enhancement \cite{lederer2017superconductivity}.

To investigate the impact of nematic quantum criticality on superconducting properties, we measured upper critical fields for superconductivity of FST thin films. We present a set of representative magnetotransport data of $H$-fixed $T$-scans (Fig. \ref{fig:Hc2}(a)) and $T$-fixed $H$-scans (Fig. \ref{fig:Hc2}(b)) for an FST thin film with $x$ = 0.13. In this study, the magnetic field is applied in the out-of-plane ($H || c$) direction. We define \tc\ and \Hc\ at the point where $\rho$($T$, $H$) becomes half of its value in the normal state (50\% criterion), as indicated by the arrows in Figs. \ref{fig:Hc2}(a) and \ref{fig:Hc2}(b). To explore the $x$-dependent evolution of the pair-breaking mechanism, we constructed $H$-$T$ phase diagrams from magnetotransport experiments on a series of FST thin films (Fig. \ref{fig:Hc2}(c)). In Se-rich samples, such as $x$ = 0.72, \Hc\ exhibits an almost linear increase upon decreasing $T$, with a relatively small initial slope at the zero-field critical temperature $T_{\rm c, 0}$, $dH/dT|_{T_{\rm c, 0}} \sim -4.65$ (T/K). In contrast, in Te-rich compositions such as $x$ = 0.06, \Hc\ shows a sharp increase with an upward convex shape near $T_{\rm c, 0}$ and a significantly larger initial slope of $dH/dT|_{T_{\rm c, 0}} \sim -15.48$ (T/K).

There are two kinds of pair-breaking effects in a type-II superconductor: the orbital effect and the Pauli paramagnetic effect. The orbital effect is associated with the kinetic energy of the supercurrent around vortices, while the Pauli effect arises from the gain of Zeeman energy by destroying spin-singlet Cooper pairing. There is a notable difference in $T$-dependence of the two critical fields near $T_{\rm c, 0}$: $H_{\rm c2}^{\rm orb.} \propto (T_{\rm c, 0} - T)$ for the orbital-limiting upper critical field, while $H_{\rm c2}^{\rm P} \propto (T_{\rm c, 0} - T)^{1/2}$ for the Pauli-limiting upper critical field. The $T$-linear initial slope for $x$ = 0.72 thus suggests a substantial orbital effect, while the $T$-sublinear dependence for $x$ = 0.06 indicates a dominant Pauli effect.

To clearly visualize the quantum critical region in the phase diagram, we constructed a color plot for the power-law exponent $n$ for $\rho$-$T$ (Fig. \ref{fig:phase}(a)). We calculated $n$ as
\begin{equation}
n = \partial\ln(\rho(T) - \rho_0)/\partial\ln{T},
\end{equation}
where residual resistivity $\rho_0$ is determined by fitting the data using the empirical model $\rho = \rho_0 + AT^n$ at low-temperature region above \tc. Remarkably, we found that the $T$-linear resistivity is observed over a wide $x$ range of 0.06 $\leq x \leq$ 0.45, which appears to reconcile with the $x$ range where the logarithmic divergence of $S/T$ is observed down to \tc. Intriguingly, as shown in Fig. \ref{fig:phase}(b), the logarithmic-term prefactor $|A_S|$ shows a pronounced peak notably at the nematic QCP $x$ = 0.45, suggesting the dominant role of nematic fluctuations in driving the NFL transport. As $x$ decreases, $|A_S|$ exhibits a monotonic decline towards the AFM QCP at $x$ = 0.08 without a peak, and is further suppressed within the AFM phase. This result suggests that the AFM QCP could be detrimental to quantum fluctuations, which appears to be consistent with the first-order phase transition in FeTe \cite{bao2009tunable, kim2023kondo} that is in general not preferable to enhancing quantum fluctuations. 

The NFL behavior gradually fades out at high temperatures above approximately 50 K, and the resistivity turns to decrease ($n <$ 0) above around 100 K. Notably, $T_{S}^{*}$ estimated from $S/T$ (Fig. \ref{fig:ST}(h)) closely aligns with $T_{\rho}^{*}$, or identically the contour $n$ = 0, indicating a good agreement in the onset temperature below which the NFL behavior begins to develop in both resistivity and thermoelectricity (Fig. \ref{fig:phase}(a)). Interestingly, the breakdown of the $T$-linear resistivity in FST at relatively low temperatures contrasts with the transport properties in cuprates, where $T$-linear resistivity persists up to 500 K \cite{phillips2022stranger}. The behavior of FST rather resembles that observed in heavy fermion systems, where itinerant electrons undergo incoherent scatterings above the Kondo temperatures. The high-temperature insulating behavior in FST might be attributed to incoherent Kondo scattering with localized Fe 3$d_{xy}$ electrons, which stems from the strong Hund coupling (orbital-selective Mott phase) \cite{kim2023kondo}, reflecting the stronger electronic correlations towards the end composition $x$ = 0 (see Supplemental Materials \cite{SM} for more details). 

A further similarity between FST and heavy fermion superconductors lies in their pair-breaking mechanisms. To quantify the dominant pair-breaking effects, it is useful to introduce the ratio of the two upper critical fields, the Maki parameter $\alpha$ = $\sqrt{2}$\Hco/\HcP, which becomes larger (smaller) than unity when the Pauli (orbital) effect is dominated. We calculated $\alpha$ by performing fittings with the Werthamer-Helfand-Hohenberg (WHH) formula, incorporating both orbital and Pauli effects \cite{WHH1966} (see Supplemental Materials \cite{SM} for more details). The fitting results are presented as the solid lines in Figs. \ref{fig:Hc2}(c) and \ref{fig:Hc2}(d). We found that $\alpha$ is negligibly small at $x$ = 0.72, but increases with decreasing $x$, and becomes well above unity in Te-rich FSTs (Fig. \ref{fig:phase}(b)). We stress that the large $\alpha$ exceeding unity for the out-of-plane field configuration is extremely rare and reported only in heavy fermion systems \cite{matsuda2007FFLO} and FST with $x$ around 0.45 \cite{Fang2010Weak, khim2010evidence, her2015anisotropy}, while it can be easily found for in-plane configuration in various low-dimensional materials \cite{nakamura2019pauli, wang2021isotropic, Agosta2017organicFFLO, cao2018unconventional, wan2023orbital} (See Supplemental Materials for in-plane \Hc\ of FST thin films \cite{SM}), where the orbital motion perpendicular to magnetic field is quenched due to dimensionality. Remarkably, we found a notable coincidence in the $x$ range (0.06 $\leq x \leq$ 0.45), where $\alpha$ significantly exceeds unity and the NFL transport behavior is observed. This coincidence can be naturally understood in terms of the mass enhancement, because $\alpha$ is also related to the effective mass $m^{*}$ in the BCS limit \cite{kasahara2014field} as 
\begin{equation}
\alpha \approx 2\dfrac{m^{*}}{m_0}\dfrac{\Delta}{E_{\rm F}},
\end{equation}
where $m_0$ is the bare electron mass, $\Delta$ is superconducting gap, and $E_{\rm F}$ is Fermi energy. Indeed, a large $\alpha$ exceeding unity in heavy fermions is predominantly driven by the strong mass renormalisation via Kondo effect \cite{matsuda2007FFLO}. Our results thus suggest a close link between the unusually Pauli-limited \Hc\ and the nematic quantum criticality in FST.

We note that the WHH fittings reasonably capture all the \Hc\ data except for low-temperature regime in $x$ = 0.06 and 0.13, where deviations from the fitting are observed as slight upturns in \Hc, indicated by the shaded areas in Figs. \ref{fig:Hc2}(c) and \ref{fig:Hc2}(d). As sufficiently large $\alpha$ ($> 1.5$) is one of the requisites for superconductors to enter the Fulde–Ferrell–Larkin–Ovchinnikov (FFLO) state \cite{matsuda2007FFLO}, it is tempting to attribute these upturns to the appearance of an FFLO state. However, stabilizing an FFLO state in such a "dirty" superconductor, where $\ell$, coherence length $\xi$, and inter-particle distance ($1/k_F$) are all comparable, is highly nontrivial, as FFLO states typically require a very clean system with $\ell \gg \xi_{ab}$. Our results thus may trigger further studies to explore exotic field-induced superconducting states driven by strong correlation and quantum fluctuations \cite{kasahara2020evidence}.

Finally, we comment on a potential existence of \textit{extended} nematic quantum criticality in FST. The NFL transport in FST is observed over an unexpectedly wide range of $x$, spanning 0.06 $\leq  x \leq$ 0.45 (Fig. \ref{fig:phase}(a)). This behavior is unusual as it differs from conventional quantum critical metals where NFL behavior is confined to the vicinity of a QCP. One intriguing interpretation of our observation is that the quantum critical regime may be stabilized across the extended parameter range, a phenomenon referred to as \textit{extended} quantum criticality \cite{phillips2022stranger}. Although such a behavior has been reported in a few materials, including cuprates \cite{jin2011link}, heavy fermions \cite{tomita2015strange}, magic-angle twisted bilayer graphene \cite{cao2020strange}, and quantum spin-liquids \cite{zhao2019quantum, wakamatsu2023thermoelectric}, it remains rare and, to the best of our knowledge, has not been observed in iron-based superconductors. Further high-field experiments are essential to probe the normal-state transport properties underneath the superconducting dome and determine whether the nematic quantum critical regime extends over a finite parameter range of $x$ in the zero-temperature limit.

In summary, we have investigated the normal-state electrical and thermoelectric transport properties of MBE-grown FST thin films across the pure nematic QCP. In the vicinity of the nematic QCP $x$ = 0.45, we find hallmark NFL transport signatures, including $T$-linear resistivity and logarithmic divergence of $S/T$, indicating that nematic fluctuations play a key role in driving NFL behavior. Additionally, our estimation of upper critical fields across the nematic QCP reveals a crossover in pair-breaking mechanisms from orbital to Pauli-limited effects with increasing Te content, highlighting the significant impact of nematic criticality-induced mass enhancement on superconducting properties. These findings underscore the profound influence of nematic fluctuations on both normal-state and superconducting behaviors.

\textbf{ACKNOWLEDGMENTS} \\
We thank N. Nagaosa, Y. Iwasa, S. Kasahara, M. Hirschberger, and R. Yamada for discussions. This work was supported by JSPS KAKENHI Grants (No. 24K17020, No. 22K18965, No. 23H04017, No. 23H05431, No. 23H05462, and No. 23H04862), JST FOREST (Grant No. JPMJFR2038), JST CREST (Grant No. JPMJCR1874 and No. JPMJCR23O3), Mitsubishi Foundation, the special fund of Institute of Industrial Science, The University of Tokyo, and the RIKEN TRIP initiative (Many-body Electron Systems).
\bibliographystyle{apsrev}

\begin{thebibliography}{10}%
\makeatletter
\providecommand \@ifxundefined [1]{%
 \@ifx{#1\undefined}
}%
\providecommand \@ifnum [1]{%
 \ifnum #1\expandafter \@firstoftwo
 \else \expandafter \@secondoftwo
 \fi
}%
\providecommand \@ifx [1]{%
 \ifx #1\expandafter \@firstoftwo
 \else \expandafter \@secondoftwo
 \fi
}%
\providecommand \natexlab [1]{#1}%
\providecommand \enquote  [1]{``#1''}%
\providecommand \bibnamefont  [1]{#1}%
\providecommand \bibfnamefont [1]{#1}%
\providecommand \citenamefont [1]{#1}%
\providecommand \href@noop [0]{\@secondoftwo}%
\providecommand \href [0]{\begingroup \@sanitize@url \@href}%
\providecommand \@href[1]{\@@startlink{#1}\@@href}%
\providecommand \@@href[1]{\endgroup#1\@@endlink}%
\providecommand \@sanitize@url [0]{\catcode `\\12\catcode `\$12\catcode
  `\&12\catcode `\#12\catcode `\^12\catcode `\_12\catcode `\%12\relax}%
\providecommand \@@startlink[1]{}%
\providecommand \@@endlink[0]{}%
\providecommand \url  [0]{\begingroup\@sanitize@url \@url }%
\providecommand \@url [1]{\endgroup\@href {#1}{\urlprefix }}%
\providecommand \urlprefix  [0]{URL }%
\providecommand \Eprint [0]{\href }%
\providecommand \doibase [0]{https://doi.org/}%
\providecommand \selectlanguage [0]{\@gobble}%
\providecommand \bibinfo  [0]{\@secondoftwo}%
\providecommand \bibfield  [0]{\@secondoftwo}%
\providecommand \translation [1]{[#1]}%
\providecommand \BibitemOpen [0]{}%
\providecommand \bibitemStop [0]{}%
\providecommand \bibitemNoStop [0]{.\EOS\space}%
\providecommand \EOS [0]{\spacefactor3000\relax}%
\providecommand \BibitemShut  [1]{\csname bibitem#1\endcsname}%
\let\auto@bib@innerbib\@empty
\bibitem [{\citenamefont {Sato}\ \emph {et~al.}(2024)\citenamefont {Sato},
  \citenamefont {Nagahama}, \citenamefont {Belopolski}, \citenamefont
  {Yoshimi}, \citenamefont {Kawamura}, \citenamefont {Tsukazaki}, \citenamefont
  {Kanazawa}, \citenamefont {Takahashi}, \citenamefont {Kawasaki},\ and\
  \citenamefont {Tokura}}]{Sato2024Molecular}%
  \BibitemOpen
  \bibfield  {author} {\bibinfo {author} {\bibfnamefont {Y.}~\bibnamefont
  {Sato}}, \bibinfo {author} {\bibfnamefont {S.}~\bibnamefont {Nagahama}},
  \bibinfo {author} {\bibfnamefont {I.}~\bibnamefont {Belopolski}}, \bibinfo
  {author} {\bibfnamefont {R.}~\bibnamefont {Yoshimi}}, \bibinfo {author}
  {\bibfnamefont {M.}~\bibnamefont {Kawamura}}, \bibinfo {author}
  {\bibfnamefont {A.}~\bibnamefont {Tsukazaki}}, \bibinfo {author}
  {\bibfnamefont {N.}~\bibnamefont {Kanazawa}}, \bibinfo {author}
  {\bibfnamefont {K.~S.}\ \bibnamefont {Takahashi}}, \bibinfo {author}
  {\bibfnamefont {M.}~\bibnamefont {Kawasaki}},\ and\ \bibinfo {author}
  {\bibfnamefont {Y.}~\bibnamefont {Tokura}},\ }\bibfield  {title} {\bibinfo
  {title} {Molecular beam epitaxy of superconducting
  $\mathrm{FeS}{\mathrm{e}}_{x}\mathrm{T}{\mathrm{e}}_{1\ensuremath{-}x}$ thin
  films interfaced with magnetic topological insulators},\ }\href@noop {}
  {\bibfield  {journal} {\bibinfo  {journal} {Phys. Rev. Mater.}\ }\textbf
  {\bibinfo {volume} {8}},\ \bibinfo {pages} {L041801} (\bibinfo {year}
  {2024})}\BibitemShut {NoStop}%
\bibitem [{\citenamefont {Mukasa}\ \emph {et~al.}(2021)\citenamefont {Mukasa},
  \citenamefont {Matsuura}, \citenamefont {Qiu}, \citenamefont {Saito},
  \citenamefont {Sugimura}, \citenamefont {Ishida}, \citenamefont {Otani},
  \citenamefont {Onishi}, \citenamefont {Mizukami}, \citenamefont {Hashimoto}
  \emph {et~al.}}]{mukasa2021high}%
  \BibitemOpen
  \bibfield  {author} {\bibinfo {author} {\bibfnamefont {K.}~\bibnamefont
  {Mukasa}}, \bibinfo {author} {\bibfnamefont {K.}~\bibnamefont {Matsuura}},
  \bibinfo {author} {\bibfnamefont {M.}~\bibnamefont {Qiu}}, \bibinfo {author}
  {\bibfnamefont {M.}~\bibnamefont {Saito}}, \bibinfo {author} {\bibfnamefont
  {Y.}~\bibnamefont {Sugimura}}, \bibinfo {author} {\bibfnamefont
  {K.}~\bibnamefont {Ishida}}, \bibinfo {author} {\bibfnamefont
  {M.}~\bibnamefont {Otani}}, \bibinfo {author} {\bibfnamefont
  {Y.}~\bibnamefont {Onishi}}, \bibinfo {author} {\bibfnamefont
  {Y.}~\bibnamefont {Mizukami}}, \bibinfo {author} {\bibfnamefont
  {K.}~\bibnamefont {Hashimoto}}, \emph {et~al.},\ }\bibfield  {title}
  {\bibinfo {title} {{High-pressure phase diagrams of FeSe$_{1-x}$Te$_x$:
  correlation between suppressed nematicity and enhanced superconductivity}},\
  }\href@noop {} {\bibfield  {journal} {\bibinfo  {journal} {Nature
  Communications}\ }\textbf {\bibinfo {volume} {12}},\ \bibinfo {pages} {381}
  (\bibinfo {year} {2021})}\BibitemShut {NoStop}%
\bibitem [{\citenamefont {Sun}\ \emph {et~al.}(2016)\citenamefont {Sun},
  \citenamefont {Yamada}, \citenamefont {Pyon},\ and\ \citenamefont
  {Tamegai}}]{sun2016influence}%
  \BibitemOpen
  \bibfield  {author} {\bibinfo {author} {\bibfnamefont {Y.}~\bibnamefont
  {Sun}}, \bibinfo {author} {\bibfnamefont {T.}~\bibnamefont {Yamada}},
  \bibinfo {author} {\bibfnamefont {S.}~\bibnamefont {Pyon}},\ and\ \bibinfo
  {author} {\bibfnamefont {T.}~\bibnamefont {Tamegai}},\ }\bibfield  {title}
  {\bibinfo {title} {{Influence of interstitial Fe to the phase diagram of
  Fe$_{1+y}$Te$_{1-x}$Se$_x$ single crystals}},\ }\href@noop {} {\bibfield
  {journal} {\bibinfo  {journal} {Scientific reports}\ }\textbf {\bibinfo
  {volume} {6}},\ \bibinfo {pages} {32290} (\bibinfo {year}
  {2016})}\BibitemShut {NoStop}%
\bibitem [{\citenamefont {Mitamura}\ \emph {et~al.}(2020)\citenamefont
  {Mitamura}, \citenamefont {Watanuki}, \citenamefont {Kampert}, \citenamefont
  {Förster}, \citenamefont {Matsuo}, \citenamefont {Onimaru}, \citenamefont
  {Onozaki}, \citenamefont {Amou}, \citenamefont {Wakiya}, \citenamefont
  {Matsumoto}, \citenamefont {Yamamoto}, \citenamefont {Suzuki}, \citenamefont
  {Zherlitsyn}, \citenamefont {Wosnitza}, \citenamefont {Tokunaga},
  \citenamefont {Kindo},\ and\ \citenamefont {Sakakibara}}]{Mitamura2020}%
  \BibitemOpen
  \bibfield  {author} {\bibinfo {author} {\bibfnamefont {H.}~\bibnamefont
  {Mitamura}}, \bibinfo {author} {\bibfnamefont {R.}~\bibnamefont {Watanuki}},
  \bibinfo {author} {\bibfnamefont {E.}~\bibnamefont {Kampert}}, \bibinfo
  {author} {\bibfnamefont {T.}~\bibnamefont {Förster}}, \bibinfo {author}
  {\bibfnamefont {A.}~\bibnamefont {Matsuo}}, \bibinfo {author} {\bibfnamefont
  {T.}~\bibnamefont {Onimaru}}, \bibinfo {author} {\bibfnamefont
  {N.}~\bibnamefont {Onozaki}}, \bibinfo {author} {\bibfnamefont
  {Y.}~\bibnamefont {Amou}}, \bibinfo {author} {\bibfnamefont {K.}~\bibnamefont
  {Wakiya}}, \bibinfo {author} {\bibfnamefont {K.~T.}\ \bibnamefont
  {Matsumoto}}, \bibinfo {author} {\bibfnamefont {I.}~\bibnamefont {Yamamoto}},
  \bibinfo {author} {\bibfnamefont {K.}~\bibnamefont {Suzuki}}, \bibinfo
  {author} {\bibfnamefont {S.}~\bibnamefont {Zherlitsyn}}, \bibinfo {author}
  {\bibfnamefont {J.}~\bibnamefont {Wosnitza}}, \bibinfo {author}
  {\bibfnamefont {M.}~\bibnamefont {Tokunaga}}, \bibinfo {author}
  {\bibfnamefont {K.}~\bibnamefont {Kindo}},\ and\ \bibinfo {author}
  {\bibfnamefont {T.}~\bibnamefont {Sakakibara}},\ }\bibfield  {title}
  {\bibinfo {title} {{Improved accuracy in high-frequency AC transport
  measurements in pulsed high magnetic fields}},\ }\href
  {https://doi.org/10.1063/5.0014986} {\bibfield  {journal} {\bibinfo
  {journal} {Review of Scientific Instruments}\ }\textbf {\bibinfo {volume}
  {91}},\ \bibinfo {pages} {125107} (\bibinfo {year} {2020})}\BibitemShut
  {NoStop}%
\bibitem [{\citenamefont {Yi}\ \emph {et~al.}(2017)\citenamefont {Yi},
  \citenamefont {Zhang}, \citenamefont {Shen},\ and\ \citenamefont
  {Lu}}]{yi2017role}%
  \BibitemOpen
  \bibfield  {author} {\bibinfo {author} {\bibfnamefont {M.}~\bibnamefont
  {Yi}}, \bibinfo {author} {\bibfnamefont {Y.}~\bibnamefont {Zhang}}, \bibinfo
  {author} {\bibfnamefont {Z.-X.}\ \bibnamefont {Shen}},\ and\ \bibinfo
  {author} {\bibfnamefont {D.}~\bibnamefont {Lu}},\ }\bibfield  {title}
  {\bibinfo {title} {Role of the orbital degree of freedom in iron-based
  superconductors},\ }\href@noop {} {\bibfield  {journal} {\bibinfo  {journal}
  {npj Quantum Materials}\ }\textbf {\bibinfo {volume} {2}},\ \bibinfo {pages}
  {57} (\bibinfo {year} {2017})}\BibitemShut {NoStop}%
\bibitem [{\citenamefont {Kim}\ \emph {et~al.}(2023)\citenamefont {Kim},
  \citenamefont {Kim}, \citenamefont {Kim}, \citenamefont {Kim}, \citenamefont
  {Kim}, \citenamefont {Cheng}, \citenamefont {Choi}, \citenamefont {Jung},
  \citenamefont {Lu}, \citenamefont {Kim} \emph {et~al.}}]{kim2023kondo}%
  \BibitemOpen
  \bibfield  {author} {\bibinfo {author} {\bibfnamefont {Y.}~\bibnamefont
  {Kim}}, \bibinfo {author} {\bibfnamefont {M.-S.}\ \bibnamefont {Kim}},
  \bibinfo {author} {\bibfnamefont {D.}~\bibnamefont {Kim}}, \bibinfo {author}
  {\bibfnamefont {M.}~\bibnamefont {Kim}}, \bibinfo {author} {\bibfnamefont
  {M.}~\bibnamefont {Kim}}, \bibinfo {author} {\bibfnamefont {C.-M.}\
  \bibnamefont {Cheng}}, \bibinfo {author} {\bibfnamefont {J.}~\bibnamefont
  {Choi}}, \bibinfo {author} {\bibfnamefont {S.}~\bibnamefont {Jung}}, \bibinfo
  {author} {\bibfnamefont {D.}~\bibnamefont {Lu}}, \bibinfo {author}
  {\bibfnamefont {J.~H.}\ \bibnamefont {Kim}}, \emph {et~al.},\ }\bibfield
  {title} {\bibinfo {title} {{Kondo interaction in FeTe and its potential role
  in the magnetic order}},\ }\href@noop {} {\bibfield  {journal} {\bibinfo
  {journal} {Nature communications}\ }\textbf {\bibinfo {volume} {14}},\
  \bibinfo {pages} {4145} (\bibinfo {year} {2023})}\BibitemShut {NoStop}%
\bibitem [{\citenamefont {Otsuka}\ \emph {et~al.}(2019)\citenamefont {Otsuka},
  \citenamefont {Hagisawa}, \citenamefont {Koshika}, \citenamefont {Adachi},
  \citenamefont {Usui}, \citenamefont {Sasaki}, \citenamefont {Sasaki},
  \citenamefont {Yamaguchi}, \citenamefont {Nakanishi}, \citenamefont
  {Yoshizawa} \emph {et~al.}}]{otsuka2019incoherent}%
  \BibitemOpen
  \bibfield  {author} {\bibinfo {author} {\bibfnamefont {T.}~\bibnamefont
  {Otsuka}}, \bibinfo {author} {\bibfnamefont {S.}~\bibnamefont {Hagisawa}},
  \bibinfo {author} {\bibfnamefont {Y.}~\bibnamefont {Koshika}}, \bibinfo
  {author} {\bibfnamefont {S.}~\bibnamefont {Adachi}}, \bibinfo {author}
  {\bibfnamefont {T.}~\bibnamefont {Usui}}, \bibinfo {author} {\bibfnamefont
  {N.}~\bibnamefont {Sasaki}}, \bibinfo {author} {\bibfnamefont
  {S.}~\bibnamefont {Sasaki}}, \bibinfo {author} {\bibfnamefont
  {S.}~\bibnamefont {Yamaguchi}}, \bibinfo {author} {\bibfnamefont
  {Y.}~\bibnamefont {Nakanishi}}, \bibinfo {author} {\bibfnamefont
  {M.}~\bibnamefont {Yoshizawa}}, \emph {et~al.},\ }\bibfield  {title}
  {\bibinfo {title} {{Incoherent-coherent crossover and the pseudogap in
  Te-annealed superconducting Fe$_{1+y}$Te$_{1-x}$Se$_x$ revealed by
  magnetotransport measurements}},\ }\href@noop {} {\bibfield  {journal}
  {\bibinfo  {journal} {Physical Review B}\ }\textbf {\bibinfo {volume} {99}},\
  \bibinfo {pages} {184505} (\bibinfo {year} {2019})}\BibitemShut {NoStop}%
\bibitem [{\citenamefont {Phillips}\ \emph {et~al.}(2022)\citenamefont
  {Phillips}, \citenamefont {Hussey},\ and\ \citenamefont
  {Abbamonte}}]{phillips2022stranger}%
  \BibitemOpen
  \bibfield  {author} {\bibinfo {author} {\bibfnamefont {P.~W.}\ \bibnamefont
  {Phillips}}, \bibinfo {author} {\bibfnamefont {N.~E.}\ \bibnamefont
  {Hussey}},\ and\ \bibinfo {author} {\bibfnamefont {P.}~\bibnamefont
  {Abbamonte}},\ }\bibfield  {title} {\bibinfo {title} {Stranger than metals},\
  }\href@noop {} {\bibfield  {journal} {\bibinfo  {journal} {Science}\ }\textbf
  {\bibinfo {volume} {377}},\ \bibinfo {pages} {eabh4273} (\bibinfo {year}
  {2022})}\BibitemShut {NoStop}%
\bibitem [{\citenamefont {Paul}\ and\ \citenamefont
  {Kotliar}(2001)}]{Paul2001Thermoelectric}%
  \BibitemOpen
  \bibfield  {author} {\bibinfo {author} {\bibfnamefont {I.}~\bibnamefont
  {Paul}}\ and\ \bibinfo {author} {\bibfnamefont {G.}~\bibnamefont {Kotliar}},\
  }\bibfield  {title} {\bibinfo {title} {Thermoelectric behavior near the
  magnetic quantum critical point},\ }\href
  {https://doi.org/10.1103/PhysRevB.64.184414} {\bibfield  {journal} {\bibinfo
  {journal} {Phys. Rev. B}\ }\textbf {\bibinfo {volume} {64}},\ \bibinfo
  {pages} {184414} (\bibinfo {year} {2001})}\BibitemShut {NoStop}%
\bibitem [{\citenamefont {Werthamer}\ \emph {et~al.}(1966)\citenamefont
  {Werthamer}, \citenamefont {Helfand},\ and\ \citenamefont
  {Hohenberg}}]{WHH1966}%
  \BibitemOpen
  \bibfield  {author} {\bibinfo {author} {\bibfnamefont {N.~R.}\ \bibnamefont
  {Werthamer}}, \bibinfo {author} {\bibfnamefont {E.}~\bibnamefont {Helfand}},\
  and\ \bibinfo {author} {\bibfnamefont {P.~C.}\ \bibnamefont {Hohenberg}},\
  }\bibfield  {title} {\bibinfo {title} {{Temperature and Purity Dependence of
  the Superconducting Critical Field, ${H}_{c2}$. III. Electron Spin and
  Spin-Orbit Effects}},\ }\href {https://doi.org/10.1103/PhysRev.147.295}
  {\bibfield  {journal} {\bibinfo  {journal} {Phys. Rev.}\ }\textbf {\bibinfo
  {volume} {147}},\ \bibinfo {pages} {295} (\bibinfo {year}
  {1966})}\BibitemShut {NoStop}%
\end{thebibliography}%


\begin{thebibliography}{51}
\expandafter\ifx\csname natexlab\endcsname\relax\def\natexlab#1{#1}\fi
\expandafter\ifx\csname bibnamefont\endcsname\relax
  \def\bibnamefont#1{#1}\fi
\expandafter\ifx\csname bibfnamefont\endcsname\relax
  \def\bibfnamefont#1{#1}\fi
\expandafter\ifx\csname citenamefont\endcsname\relax
  \def\citenamefont#1{#1}\fi
\expandafter\ifx\csname url\endcsname\relax
  \def\url#1{\texttt{#1}}\fi
\expandafter\ifx\csname urlprefix\endcsname\relax\def\urlprefix{URL }\fi
\providecommand{\bibinfo}[2]{#2}
\providecommand{\eprint}[2][]{\url{#2}}

\bibitem[{\citenamefont{L\"ohneysen et~al.}(2007)\citenamefont{L\"ohneysen,
  Rosch, Vojta, and W\"olfle}}]{Lohneysen2007Fermi}
\bibinfo{author}{\bibfnamefont{H.~v.} \bibnamefont{L\"ohneysen}},
  \bibinfo{author}{\bibfnamefont{A.}~\bibnamefont{Rosch}},
  \bibinfo{author}{\bibfnamefont{M.}~\bibnamefont{Vojta}}, \bibnamefont{and}
  \bibinfo{author}{\bibfnamefont{P.}~\bibnamefont{W\"olfle}},
  \bibinfo{journal}{Rev. Mod. Phys.} \textbf{\bibinfo{volume}{79}},
  \bibinfo{pages}{1015} (\bibinfo{year}{2007}).

\bibitem[{\citenamefont{Shibauchi et~al.}(2014)\citenamefont{Shibauchi,
  Carrington, and Matsuda}}]{shibauchi2014quantum}
\bibinfo{author}{\bibfnamefont{T.}~\bibnamefont{Shibauchi}},
  \bibinfo{author}{\bibfnamefont{A.}~\bibnamefont{Carrington}},
  \bibnamefont{and} \bibinfo{author}{\bibfnamefont{Y.}~\bibnamefont{Matsuda}},
  \bibinfo{journal}{Annu. Rev. Condens. Matter Phys.}
  \textbf{\bibinfo{volume}{5}}, \bibinfo{pages}{113} (\bibinfo{year}{2014}).

\bibitem[{\citenamefont{Paschen and Si}(2021)}]{paschen2021quantum}
\bibinfo{author}{\bibfnamefont{S.}~\bibnamefont{Paschen}} \bibnamefont{and}
  \bibinfo{author}{\bibfnamefont{Q.}~\bibnamefont{Si}},
  \bibinfo{journal}{Nature Reviews Physics} \textbf{\bibinfo{volume}{3}},
  \bibinfo{pages}{9} (\bibinfo{year}{2021}).

\bibitem[{\citenamefont{Phillips et~al.}(2022)\citenamefont{Phillips, Hussey,
  and Abbamonte}}]{phillips2022stranger}
\bibinfo{author}{\bibfnamefont{P.~W.} \bibnamefont{Phillips}},
  \bibinfo{author}{\bibfnamefont{N.~E.} \bibnamefont{Hussey}},
  \bibnamefont{and}
  \bibinfo{author}{\bibfnamefont{P.}~\bibnamefont{Abbamonte}},
  \bibinfo{journal}{Science} \textbf{\bibinfo{volume}{377}},
  \bibinfo{pages}{eabh4273} (\bibinfo{year}{2022}).

\bibitem[{\citenamefont{Sato et~al.}(2017)\citenamefont{Sato, Kasahara,
  Murayama, Kasahara, Moon, Nishizaki, Loew, Porras, Keimer, Shibauchi
  et~al.}}]{sato2017thermodynamic}
\bibinfo{author}{\bibfnamefont{Y.}~\bibnamefont{Sato}},
  \bibinfo{author}{\bibfnamefont{S.}~\bibnamefont{Kasahara}},
  \bibinfo{author}{\bibfnamefont{H.}~\bibnamefont{Murayama}},
  \bibinfo{author}{\bibfnamefont{Y.}~\bibnamefont{Kasahara}},
  \bibinfo{author}{\bibfnamefont{E.-G.} \bibnamefont{Moon}},
  \bibinfo{author}{\bibfnamefont{T.}~\bibnamefont{Nishizaki}},
  \bibinfo{author}{\bibfnamefont{T.}~\bibnamefont{Loew}},
  \bibinfo{author}{\bibfnamefont{J.}~\bibnamefont{Porras}},
  \bibinfo{author}{\bibfnamefont{B.}~\bibnamefont{Keimer}},
  \bibinfo{author}{\bibfnamefont{T.}~\bibnamefont{Shibauchi}},
  \bibnamefont{et~al.}, \bibinfo{journal}{Nature Physics}
  \textbf{\bibinfo{volume}{13}}, \bibinfo{pages}{1074} (\bibinfo{year}{2017}).

\bibitem[{\citenamefont{Fernandes et~al.}(2022)\citenamefont{Fernandes, Coldea,
  Ding, Fisher, Hirschfeld, and Kotliar}}]{fernandes2022iron}
\bibinfo{author}{\bibfnamefont{R.~M.} \bibnamefont{Fernandes}},
  \bibinfo{author}{\bibfnamefont{A.~I.} \bibnamefont{Coldea}},
  \bibinfo{author}{\bibfnamefont{H.}~\bibnamefont{Ding}},
  \bibinfo{author}{\bibfnamefont{I.~R.} \bibnamefont{Fisher}},
  \bibinfo{author}{\bibfnamefont{P.}~\bibnamefont{Hirschfeld}},
  \bibnamefont{and} \bibinfo{author}{\bibfnamefont{G.}~\bibnamefont{Kotliar}},
  \bibinfo{journal}{Nature} \textbf{\bibinfo{volume}{601}}, \bibinfo{pages}{35}
  (\bibinfo{year}{2022}).

\bibitem[{\citenamefont{Nie et~al.}(2022)\citenamefont{Nie, Sun, Ma, Song,
  Zheng, Liang, Wu, Yu, Li, Shan et~al.}}]{nie2022charge}
\bibinfo{author}{\bibfnamefont{L.}~\bibnamefont{Nie}},
  \bibinfo{author}{\bibfnamefont{K.}~\bibnamefont{Sun}},
  \bibinfo{author}{\bibfnamefont{W.}~\bibnamefont{Ma}},
  \bibinfo{author}{\bibfnamefont{D.}~\bibnamefont{Song}},
  \bibinfo{author}{\bibfnamefont{L.}~\bibnamefont{Zheng}},
  \bibinfo{author}{\bibfnamefont{Z.}~\bibnamefont{Liang}},
  \bibinfo{author}{\bibfnamefont{P.}~\bibnamefont{Wu}},
  \bibinfo{author}{\bibfnamefont{F.}~\bibnamefont{Yu}},
  \bibinfo{author}{\bibfnamefont{J.}~\bibnamefont{Li}},
  \bibinfo{author}{\bibfnamefont{M.}~\bibnamefont{Shan}}, \bibnamefont{et~al.},
  \bibinfo{journal}{Nature} \textbf{\bibinfo{volume}{604}}, \bibinfo{pages}{59}
  (\bibinfo{year}{2022}).

\bibitem[{\citenamefont{Cao et~al.}(2021)\citenamefont{Cao, Rodan-Legrain,
  Park, Yuan, Watanabe, Taniguchi, Fernandes, Fu, and
  Jarillo-Herrero}}]{cao2021nematicity}
\bibinfo{author}{\bibfnamefont{Y.}~\bibnamefont{Cao}},
  \bibinfo{author}{\bibfnamefont{D.}~\bibnamefont{Rodan-Legrain}},
  \bibinfo{author}{\bibfnamefont{J.~M.} \bibnamefont{Park}},
  \bibinfo{author}{\bibfnamefont{N.~F.} \bibnamefont{Yuan}},
  \bibinfo{author}{\bibfnamefont{K.}~\bibnamefont{Watanabe}},
  \bibinfo{author}{\bibfnamefont{T.}~\bibnamefont{Taniguchi}},
  \bibinfo{author}{\bibfnamefont{R.~M.} \bibnamefont{Fernandes}},
  \bibinfo{author}{\bibfnamefont{L.}~\bibnamefont{Fu}}, \bibnamefont{and}
  \bibinfo{author}{\bibfnamefont{P.}~\bibnamefont{Jarillo-Herrero}},
  \bibinfo{journal}{science} \textbf{\bibinfo{volume}{372}},
  \bibinfo{pages}{264} (\bibinfo{year}{2021}).

\bibitem[{\citenamefont{Ramshaw et~al.}(2015)\citenamefont{Ramshaw, Sebastian,
  McDonald, Day, Tan, Zhu, Betts, Liang, Bonn, Hardy
  et~al.}}]{Ramshaw2015Quasiparticle}
\bibinfo{author}{\bibfnamefont{B.~J.} \bibnamefont{Ramshaw}},
  \bibinfo{author}{\bibfnamefont{S.~E.} \bibnamefont{Sebastian}},
  \bibinfo{author}{\bibfnamefont{R.~D.} \bibnamefont{McDonald}},
  \bibinfo{author}{\bibfnamefont{J.}~\bibnamefont{Day}},
  \bibinfo{author}{\bibfnamefont{B.~S.} \bibnamefont{Tan}},
  \bibinfo{author}{\bibfnamefont{Z.}~\bibnamefont{Zhu}},
  \bibinfo{author}{\bibfnamefont{J.~B.} \bibnamefont{Betts}},
  \bibinfo{author}{\bibfnamefont{R.}~\bibnamefont{Liang}},
  \bibinfo{author}{\bibfnamefont{D.~A.} \bibnamefont{Bonn}},
  \bibinfo{author}{\bibfnamefont{W.~N.} \bibnamefont{Hardy}},
  \bibnamefont{et~al.}, \bibinfo{journal}{Science}
  \textbf{\bibinfo{volume}{348}}, \bibinfo{pages}{317} (\bibinfo{year}{2015}).

\bibitem[{\citenamefont{Shibauchi et~al.}(2020)\citenamefont{Shibauchi,
  Hanaguri, and Matsuda}}]{shibauchi2020exotic}
\bibinfo{author}{\bibfnamefont{T.}~\bibnamefont{Shibauchi}},
  \bibinfo{author}{\bibfnamefont{T.}~\bibnamefont{Hanaguri}}, \bibnamefont{and}
  \bibinfo{author}{\bibfnamefont{Y.}~\bibnamefont{Matsuda}},
  \bibinfo{journal}{Journal of the Physical Society of Japan}
  \textbf{\bibinfo{volume}{89}}, \bibinfo{pages}{102002}
  (\bibinfo{year}{2020}).

\bibitem[{\citenamefont{Yin et~al.}(2011)\citenamefont{Yin, Haule, and
  Kotliar}}]{yin2011kinetic}
\bibinfo{author}{\bibfnamefont{Z.}~\bibnamefont{Yin}},
  \bibinfo{author}{\bibfnamefont{K.}~\bibnamefont{Haule}}, \bibnamefont{and}
  \bibinfo{author}{\bibfnamefont{G.}~\bibnamefont{Kotliar}},
  \bibinfo{journal}{Nature materials} \textbf{\bibinfo{volume}{10}},
  \bibinfo{pages}{932} (\bibinfo{year}{2011}).

\bibitem[{\citenamefont{Yi et~al.}(2017)\citenamefont{Yi, Zhang, Shen, and
  Lu}}]{yi2017role}
\bibinfo{author}{\bibfnamefont{M.}~\bibnamefont{Yi}},
  \bibinfo{author}{\bibfnamefont{Y.}~\bibnamefont{Zhang}},
  \bibinfo{author}{\bibfnamefont{Z.-X.} \bibnamefont{Shen}}, \bibnamefont{and}
  \bibinfo{author}{\bibfnamefont{D.}~\bibnamefont{Lu}}, \bibinfo{journal}{npj
  Quantum Materials} \textbf{\bibinfo{volume}{2}}, \bibinfo{pages}{57}
  (\bibinfo{year}{2017}).

\bibitem[{\citenamefont{Mukasa et~al.}(2021)\citenamefont{Mukasa, Matsuura,
  Qiu, Saito, Sugimura, Ishida, Otani, Onishi, Mizukami, Hashimoto
  et~al.}}]{mukasa2021high}
\bibinfo{author}{\bibfnamefont{K.}~\bibnamefont{Mukasa}},
  \bibinfo{author}{\bibfnamefont{K.}~\bibnamefont{Matsuura}},
  \bibinfo{author}{\bibfnamefont{M.}~\bibnamefont{Qiu}},
  \bibinfo{author}{\bibfnamefont{M.}~\bibnamefont{Saito}},
  \bibinfo{author}{\bibfnamefont{Y.}~\bibnamefont{Sugimura}},
  \bibinfo{author}{\bibfnamefont{K.}~\bibnamefont{Ishida}},
  \bibinfo{author}{\bibfnamefont{M.}~\bibnamefont{Otani}},
  \bibinfo{author}{\bibfnamefont{Y.}~\bibnamefont{Onishi}},
  \bibinfo{author}{\bibfnamefont{Y.}~\bibnamefont{Mizukami}},
  \bibinfo{author}{\bibfnamefont{K.}~\bibnamefont{Hashimoto}},
  \bibnamefont{et~al.}, \bibinfo{journal}{Nature Communications}
  \textbf{\bibinfo{volume}{12}}, \bibinfo{pages}{381} (\bibinfo{year}{2021}).

\bibitem[{\citenamefont{Ishida et~al.}(2022)\citenamefont{Ishida, Onishi,
  Tsujii, Mukasa, Qiu, Saito, Sugimura, Matsuura, Mizukami, Hashimoto
  et~al.}}]{ishida2022pure}
\bibinfo{author}{\bibfnamefont{K.}~\bibnamefont{Ishida}},
  \bibinfo{author}{\bibfnamefont{Y.}~\bibnamefont{Onishi}},
  \bibinfo{author}{\bibfnamefont{M.}~\bibnamefont{Tsujii}},
  \bibinfo{author}{\bibfnamefont{K.}~\bibnamefont{Mukasa}},
  \bibinfo{author}{\bibfnamefont{M.}~\bibnamefont{Qiu}},
  \bibinfo{author}{\bibfnamefont{M.}~\bibnamefont{Saito}},
  \bibinfo{author}{\bibfnamefont{Y.}~\bibnamefont{Sugimura}},
  \bibinfo{author}{\bibfnamefont{K.}~\bibnamefont{Matsuura}},
  \bibinfo{author}{\bibfnamefont{Y.}~\bibnamefont{Mizukami}},
  \bibinfo{author}{\bibfnamefont{K.}~\bibnamefont{Hashimoto}},
  \bibnamefont{et~al.}, \bibinfo{journal}{Proceedings of the National Academy
  of Sciences} \textbf{\bibinfo{volume}{119}}, \bibinfo{pages}{e2110501119}
  (\bibinfo{year}{2022}).

\bibitem[{\citenamefont{Jiang et~al.}(2023)\citenamefont{Jiang, Shi,
  Christensen, Sanchez, Huang, Lin, Liu, Malinowski, Xu, Fernandes
  et~al.}}]{jiang2023nematic}
\bibinfo{author}{\bibfnamefont{Q.}~\bibnamefont{Jiang}},
  \bibinfo{author}{\bibfnamefont{Y.}~\bibnamefont{Shi}},
  \bibinfo{author}{\bibfnamefont{M.~H.} \bibnamefont{Christensen}},
  \bibinfo{author}{\bibfnamefont{J.~J.} \bibnamefont{Sanchez}},
  \bibinfo{author}{\bibfnamefont{B.}~\bibnamefont{Huang}},
  \bibinfo{author}{\bibfnamefont{Z.}~\bibnamefont{Lin}},
  \bibinfo{author}{\bibfnamefont{Z.}~\bibnamefont{Liu}},
  \bibinfo{author}{\bibfnamefont{P.}~\bibnamefont{Malinowski}},
  \bibinfo{author}{\bibfnamefont{X.}~\bibnamefont{Xu}},
  \bibinfo{author}{\bibfnamefont{R.~M.} \bibnamefont{Fernandes}},
  \bibnamefont{et~al.}, \bibinfo{journal}{Communications Physics}
  \textbf{\bibinfo{volume}{6}}, \bibinfo{pages}{39} (\bibinfo{year}{2023}).

\bibitem[{\citenamefont{Bao et~al.}(2009)\citenamefont{Bao, Qiu, Huang, Green,
  Zajdel, Fitzsimmons, Zhernenkov, Chang, Fang, Qian et~al.}}]{bao2009tunable}
\bibinfo{author}{\bibfnamefont{W.}~\bibnamefont{Bao}},
  \bibinfo{author}{\bibfnamefont{Y.}~\bibnamefont{Qiu}},
  \bibinfo{author}{\bibfnamefont{Q.}~\bibnamefont{Huang}},
  \bibinfo{author}{\bibfnamefont{M.}~\bibnamefont{Green}},
  \bibinfo{author}{\bibfnamefont{P.}~\bibnamefont{Zajdel}},
  \bibinfo{author}{\bibfnamefont{M.}~\bibnamefont{Fitzsimmons}},
  \bibinfo{author}{\bibfnamefont{M.}~\bibnamefont{Zhernenkov}},
  \bibinfo{author}{\bibfnamefont{S.}~\bibnamefont{Chang}},
  \bibinfo{author}{\bibfnamefont{M.}~\bibnamefont{Fang}},
  \bibinfo{author}{\bibfnamefont{B.}~\bibnamefont{Qian}}, \bibnamefont{et~al.},
  \bibinfo{journal}{Physical Review Letters} \textbf{\bibinfo{volume}{102}},
  \bibinfo{pages}{247001} (\bibinfo{year}{2009}).

\bibitem[{\citenamefont{Otsuka et~al.}(2019)\citenamefont{Otsuka, Hagisawa,
  Koshika, Adachi, Usui, Sasaki, Sasaki, Yamaguchi, Nakanishi, Yoshizawa
  et~al.}}]{otsuka2019incoherent}
\bibinfo{author}{\bibfnamefont{T.}~\bibnamefont{Otsuka}},
  \bibinfo{author}{\bibfnamefont{S.}~\bibnamefont{Hagisawa}},
  \bibinfo{author}{\bibfnamefont{Y.}~\bibnamefont{Koshika}},
  \bibinfo{author}{\bibfnamefont{S.}~\bibnamefont{Adachi}},
  \bibinfo{author}{\bibfnamefont{T.}~\bibnamefont{Usui}},
  \bibinfo{author}{\bibfnamefont{N.}~\bibnamefont{Sasaki}},
  \bibinfo{author}{\bibfnamefont{S.}~\bibnamefont{Sasaki}},
  \bibinfo{author}{\bibfnamefont{S.}~\bibnamefont{Yamaguchi}},
  \bibinfo{author}{\bibfnamefont{Y.}~\bibnamefont{Nakanishi}},
  \bibinfo{author}{\bibfnamefont{M.}~\bibnamefont{Yoshizawa}},
  \bibnamefont{et~al.}, \bibinfo{journal}{Physical Review B}
  \textbf{\bibinfo{volume}{99}}, \bibinfo{pages}{184505}
  (\bibinfo{year}{2019}).

\bibitem[{\citenamefont{Ar{\v{c}}on et~al.}(2010)\citenamefont{Ar{\v{c}}on,
  Jegli{\v{c}}, Zorko, Poto{\v{c}}nik, Ganin, Takabayashi, Rosseinsky, and
  Prassides}}]{arvcon2010coexistence}
\bibinfo{author}{\bibfnamefont{D.}~\bibnamefont{Ar{\v{c}}on}},
  \bibinfo{author}{\bibfnamefont{P.}~\bibnamefont{Jegli{\v{c}}}},
  \bibinfo{author}{\bibfnamefont{A.}~\bibnamefont{Zorko}},
  \bibinfo{author}{\bibfnamefont{A.}~\bibnamefont{Poto{\v{c}}nik}},
  \bibinfo{author}{\bibfnamefont{A.}~\bibnamefont{Ganin}},
  \bibinfo{author}{\bibfnamefont{Y.}~\bibnamefont{Takabayashi}},
  \bibinfo{author}{\bibfnamefont{M.}~\bibnamefont{Rosseinsky}},
  \bibnamefont{and}
  \bibinfo{author}{\bibfnamefont{K.}~\bibnamefont{Prassides}},
  \bibinfo{journal}{Physical Review B} \textbf{\bibinfo{volume}{82}},
  \bibinfo{pages}{140508} (\bibinfo{year}{2010}).

\bibitem[{\citenamefont{Licciardello et~al.}(2019)\citenamefont{Licciardello,
  Buhot, Lu, Ayres, Kasahara, Matsuda, Shibauchi, and
  Hussey}}]{licciardello2019electrical}
\bibinfo{author}{\bibfnamefont{S.}~\bibnamefont{Licciardello}},
  \bibinfo{author}{\bibfnamefont{J.}~\bibnamefont{Buhot}},
  \bibinfo{author}{\bibfnamefont{J.}~\bibnamefont{Lu}},
  \bibinfo{author}{\bibfnamefont{J.}~\bibnamefont{Ayres}},
  \bibinfo{author}{\bibfnamefont{S.}~\bibnamefont{Kasahara}},
  \bibinfo{author}{\bibfnamefont{Y.}~\bibnamefont{Matsuda}},
  \bibinfo{author}{\bibfnamefont{T.}~\bibnamefont{Shibauchi}},
  \bibnamefont{and} \bibinfo{author}{\bibfnamefont{N.}~\bibnamefont{Hussey}},
  \bibinfo{journal}{Nature} \textbf{\bibinfo{volume}{567}},
  \bibinfo{pages}{213} (\bibinfo{year}{2019}).

\bibitem[{\citenamefont{Lederer et~al.}(2017)\citenamefont{Lederer, Schattner,
  Berg, and Kivelson}}]{lederer2017superconductivity}
\bibinfo{author}{\bibfnamefont{S.}~\bibnamefont{Lederer}},
  \bibinfo{author}{\bibfnamefont{Y.}~\bibnamefont{Schattner}},
  \bibinfo{author}{\bibfnamefont{E.}~\bibnamefont{Berg}}, \bibnamefont{and}
  \bibinfo{author}{\bibfnamefont{S.~A.} \bibnamefont{Kivelson}},
  \bibinfo{journal}{Proceedings of the National Academy of Sciences}
  \textbf{\bibinfo{volume}{114}}, \bibinfo{pages}{4905} (\bibinfo{year}{2017}).

\bibitem[{\citenamefont{Coldea et~al.}(2019)\citenamefont{Coldea, Blake,
  Kasahara, Haghighirad, Watson, Knafo, Choi, McCollam, Reiss, Yamashita
  et~al.}}]{coldea2019evolution}
\bibinfo{author}{\bibfnamefont{A.~I.} \bibnamefont{Coldea}},
  \bibinfo{author}{\bibfnamefont{S.~F.} \bibnamefont{Blake}},
  \bibinfo{author}{\bibfnamefont{S.}~\bibnamefont{Kasahara}},
  \bibinfo{author}{\bibfnamefont{A.~A.} \bibnamefont{Haghighirad}},
  \bibinfo{author}{\bibfnamefont{M.~D.} \bibnamefont{Watson}},
  \bibinfo{author}{\bibfnamefont{W.}~\bibnamefont{Knafo}},
  \bibinfo{author}{\bibfnamefont{E.~S.} \bibnamefont{Choi}},
  \bibinfo{author}{\bibfnamefont{A.}~\bibnamefont{McCollam}},
  \bibinfo{author}{\bibfnamefont{P.}~\bibnamefont{Reiss}},
  \bibinfo{author}{\bibfnamefont{T.}~\bibnamefont{Yamashita}},
  \bibnamefont{et~al.}, \bibinfo{journal}{npj Quantum Materials}
  \textbf{\bibinfo{volume}{4}}, \bibinfo{pages}{2} (\bibinfo{year}{2019}).

\bibitem[{\citenamefont{Mukasa et~al.}(2023)\citenamefont{Mukasa, Ishida,
  Imajo, Qiu, Saito, Matsuura, Sugimura, Liu, Uezono, Otsuka
  et~al.}}]{mukasa2023enhanced}
\bibinfo{author}{\bibfnamefont{K.}~\bibnamefont{Mukasa}},
  \bibinfo{author}{\bibfnamefont{K.}~\bibnamefont{Ishida}},
  \bibinfo{author}{\bibfnamefont{S.}~\bibnamefont{Imajo}},
  \bibinfo{author}{\bibfnamefont{M.}~\bibnamefont{Qiu}},
  \bibinfo{author}{\bibfnamefont{M.}~\bibnamefont{Saito}},
  \bibinfo{author}{\bibfnamefont{K.}~\bibnamefont{Matsuura}},
  \bibinfo{author}{\bibfnamefont{Y.}~\bibnamefont{Sugimura}},
  \bibinfo{author}{\bibfnamefont{S.}~\bibnamefont{Liu}},
  \bibinfo{author}{\bibfnamefont{Y.}~\bibnamefont{Uezono}},
  \bibinfo{author}{\bibfnamefont{T.}~\bibnamefont{Otsuka}},
  \bibnamefont{et~al.}, \bibinfo{journal}{Physical Review X}
  \textbf{\bibinfo{volume}{13}}, \bibinfo{pages}{011032}
  (\bibinfo{year}{2023}).

\bibitem[{\citenamefont{Sun et~al.}(2016)\citenamefont{Sun, Yamada, Pyon, and
  Tamegai}}]{sun2016influence}
\bibinfo{author}{\bibfnamefont{Y.}~\bibnamefont{Sun}},
  \bibinfo{author}{\bibfnamefont{T.}~\bibnamefont{Yamada}},
  \bibinfo{author}{\bibfnamefont{S.}~\bibnamefont{Pyon}}, \bibnamefont{and}
  \bibinfo{author}{\bibfnamefont{T.}~\bibnamefont{Tamegai}},
  \bibinfo{journal}{Scientific reports} \textbf{\bibinfo{volume}{6}},
  \bibinfo{pages}{32290} (\bibinfo{year}{2016}).

\bibitem[{\citenamefont{Sato et~al.}(2024)\citenamefont{Sato, Nagahama,
  Belopolski, Yoshimi, Kawamura, Tsukazaki, Kanazawa, Takahashi, Kawasaki, and
  Tokura}}]{Sato2024Molecular}
\bibinfo{author}{\bibfnamefont{Y.}~\bibnamefont{Sato}},
  \bibinfo{author}{\bibfnamefont{S.}~\bibnamefont{Nagahama}},
  \bibinfo{author}{\bibfnamefont{I.}~\bibnamefont{Belopolski}},
  \bibinfo{author}{\bibfnamefont{R.}~\bibnamefont{Yoshimi}},
  \bibinfo{author}{\bibfnamefont{M.}~\bibnamefont{Kawamura}},
  \bibinfo{author}{\bibfnamefont{A.}~\bibnamefont{Tsukazaki}},
  \bibinfo{author}{\bibfnamefont{N.}~\bibnamefont{Kanazawa}},
  \bibinfo{author}{\bibfnamefont{K.~S.} \bibnamefont{Takahashi}},
  \bibinfo{author}{\bibfnamefont{M.}~\bibnamefont{Kawasaki}}, \bibnamefont{and}
  \bibinfo{author}{\bibfnamefont{Y.}~\bibnamefont{Tokura}},
  \bibinfo{journal}{Phys. Rev. Mater.} \textbf{\bibinfo{volume}{8}},
  \bibinfo{pages}{L041801} (\bibinfo{year}{2024}).

\bibitem[{\citenamefont{Legros et~al.}(2019)\citenamefont{Legros, Benhabib,
  Tabis, Lalibert{\'e}, Dion, Lizaire, Vignolle, Vignolles, Raffy, Li
  et~al.}}]{legros2019universal}
\bibinfo{author}{\bibfnamefont{A.}~\bibnamefont{Legros}},
  \bibinfo{author}{\bibfnamefont{S.}~\bibnamefont{Benhabib}},
  \bibinfo{author}{\bibfnamefont{W.}~\bibnamefont{Tabis}},
  \bibinfo{author}{\bibfnamefont{F.}~\bibnamefont{Lalibert{\'e}}},
  \bibinfo{author}{\bibfnamefont{M.}~\bibnamefont{Dion}},
  \bibinfo{author}{\bibfnamefont{M.}~\bibnamefont{Lizaire}},
  \bibinfo{author}{\bibfnamefont{B.}~\bibnamefont{Vignolle}},
  \bibinfo{author}{\bibfnamefont{D.}~\bibnamefont{Vignolles}},
  \bibinfo{author}{\bibfnamefont{H.}~\bibnamefont{Raffy}},
  \bibinfo{author}{\bibfnamefont{Z.}~\bibnamefont{Li}}, \bibnamefont{et~al.},
  \bibinfo{journal}{Nature Physics} \textbf{\bibinfo{volume}{15}},
  \bibinfo{pages}{142} (\bibinfo{year}{2019}).

\bibitem[{SM()}]{SM}
\bibinfo{note}{See Supplemental Material for details of sample fabrication,
  transport and thermoelectricity measurements, and discussion about Kondo
  effect, MIR limit, and WHH fittings.}

\bibitem[{\citenamefont{Paul and Kotliar}(2001)}]{Paul2001Thermoelectric}
\bibinfo{author}{\bibfnamefont{I.}~\bibnamefont{Paul}} \bibnamefont{and}
  \bibinfo{author}{\bibfnamefont{G.}~\bibnamefont{Kotliar}},
  \bibinfo{journal}{Phys. Rev. B} \textbf{\bibinfo{volume}{64}},
  \bibinfo{pages}{184414} (\bibinfo{year}{2001}).

\bibitem[{\citenamefont{Pallecchi et~al.}(2016)\citenamefont{Pallecchi,
  Caglieris, and Putti}}]{pallecchi2016thermoelectric}
\bibinfo{author}{\bibfnamefont{I.}~\bibnamefont{Pallecchi}},
  \bibinfo{author}{\bibfnamefont{F.}~\bibnamefont{Caglieris}},
  \bibnamefont{and} \bibinfo{author}{\bibfnamefont{M.}~\bibnamefont{Putti}},
  \bibinfo{journal}{Superconductor Science and Technology}
  \textbf{\bibinfo{volume}{29}}, \bibinfo{pages}{073002}
  (\bibinfo{year}{2016}).

\bibitem[{\citenamefont{Xu et~al.}(2023)\citenamefont{Xu, Qin, Lin, Zhang,
  Zhang, Xu, Zhang, Shi, Yuan, Zhu et~al.}}]{Xu2023In}
\bibinfo{author}{\bibfnamefont{J.}~\bibnamefont{Xu}},
  \bibinfo{author}{\bibfnamefont{M.}~\bibnamefont{Qin}},
  \bibinfo{author}{\bibfnamefont{Z.}~\bibnamefont{Lin}},
  \bibinfo{author}{\bibfnamefont{X.}~\bibnamefont{Zhang}},
  \bibinfo{author}{\bibfnamefont{R.}~\bibnamefont{Zhang}},
  \bibinfo{author}{\bibfnamefont{L.}~\bibnamefont{Xu}},
  \bibinfo{author}{\bibfnamefont{L.}~\bibnamefont{Zhang}},
  \bibinfo{author}{\bibfnamefont{Q.}~\bibnamefont{Shi}},
  \bibinfo{author}{\bibfnamefont{J.}~\bibnamefont{Yuan}},
  \bibinfo{author}{\bibfnamefont{B.}~\bibnamefont{Zhu}}, \bibnamefont{et~al.},
  \bibinfo{journal}{Phys. Rev. B} \textbf{\bibinfo{volume}{107}},
  \bibinfo{pages}{094514} (\bibinfo{year}{2023}).

\bibitem[{\citenamefont{Daou et~al.}(2009)\citenamefont{Daou, Cyr-Choiniere,
  Lalibert{\'e}, LeBoeuf, Doiron-Leyraud, Yan, Zhou, Goodenough, and
  Taillefer}}]{daou2009thermopower}
\bibinfo{author}{\bibfnamefont{R.}~\bibnamefont{Daou}},
  \bibinfo{author}{\bibfnamefont{O.}~\bibnamefont{Cyr-Choiniere}},
  \bibinfo{author}{\bibfnamefont{F.}~\bibnamefont{Lalibert{\'e}}},
  \bibinfo{author}{\bibfnamefont{D.}~\bibnamefont{LeBoeuf}},
  \bibinfo{author}{\bibfnamefont{N.}~\bibnamefont{Doiron-Leyraud}},
  \bibinfo{author}{\bibfnamefont{J.-Q.} \bibnamefont{Yan}},
  \bibinfo{author}{\bibfnamefont{J.-S.} \bibnamefont{Zhou}},
  \bibinfo{author}{\bibfnamefont{J.}~\bibnamefont{Goodenough}},
  \bibnamefont{and}
  \bibinfo{author}{\bibfnamefont{L.}~\bibnamefont{Taillefer}},
  \bibinfo{journal}{Physical Review B} \textbf{\bibinfo{volume}{79}},
  \bibinfo{pages}{180505} (\bibinfo{year}{2009}).

\bibitem[{\citenamefont{Mandal et~al.}(2019)\citenamefont{Mandal, Sarkar, and
  Greene}}]{Mandal2019anomalous}
\bibinfo{author}{\bibfnamefont{P.~R.} \bibnamefont{Mandal}},
  \bibinfo{author}{\bibfnamefont{T.}~\bibnamefont{Sarkar}}, \bibnamefont{and}
  \bibinfo{author}{\bibfnamefont{R.~L.} \bibnamefont{Greene}},
  \bibinfo{journal}{Proceedings of the National Academy of Sciences}
  \textbf{\bibinfo{volume}{116}}, \bibinfo{pages}{5991} (\bibinfo{year}{2019}).

\bibitem[{\citenamefont{Arsenijevi{\'c}
  et~al.}(2013)\citenamefont{Arsenijevi{\'c}, Hodovanets, Ga{\'a}l, Forr{\'o},
  Bud'Ko, and Canfield}}]{arsenijevic2013signatures}
\bibinfo{author}{\bibfnamefont{S.}~\bibnamefont{Arsenijevi{\'c}}},
  \bibinfo{author}{\bibfnamefont{H.}~\bibnamefont{Hodovanets}},
  \bibinfo{author}{\bibfnamefont{R.}~\bibnamefont{Ga{\'a}l}},
  \bibinfo{author}{\bibfnamefont{L.}~\bibnamefont{Forr{\'o}}},
  \bibinfo{author}{\bibfnamefont{S.}~\bibnamefont{Bud'Ko}}, \bibnamefont{and}
  \bibinfo{author}{\bibfnamefont{P.}~\bibnamefont{Canfield}},
  \bibinfo{journal}{Physical Review B―Condensed Matter and Materials Physics}
  \textbf{\bibinfo{volume}{87}}, \bibinfo{pages}{224508}
  (\bibinfo{year}{2013}).

\bibitem[{\citenamefont{Kim et~al.}(2023)\citenamefont{Kim, Kim, Kim, Kim, Kim,
  Cheng, Choi, Jung, Lu, Kim et~al.}}]{kim2023kondo}
\bibinfo{author}{\bibfnamefont{Y.}~\bibnamefont{Kim}},
  \bibinfo{author}{\bibfnamefont{M.-S.} \bibnamefont{Kim}},
  \bibinfo{author}{\bibfnamefont{D.}~\bibnamefont{Kim}},
  \bibinfo{author}{\bibfnamefont{M.}~\bibnamefont{Kim}},
  \bibinfo{author}{\bibfnamefont{M.}~\bibnamefont{Kim}},
  \bibinfo{author}{\bibfnamefont{C.-M.} \bibnamefont{Cheng}},
  \bibinfo{author}{\bibfnamefont{J.}~\bibnamefont{Choi}},
  \bibinfo{author}{\bibfnamefont{S.}~\bibnamefont{Jung}},
  \bibinfo{author}{\bibfnamefont{D.}~\bibnamefont{Lu}},
  \bibinfo{author}{\bibfnamefont{J.~H.} \bibnamefont{Kim}},
  \bibnamefont{et~al.}, \bibinfo{journal}{Nature communications}
  \textbf{\bibinfo{volume}{14}}, \bibinfo{pages}{4145} (\bibinfo{year}{2023}).

\bibitem[{\citenamefont{Werthamer et~al.}(1966)\citenamefont{Werthamer,
  Helfand, and Hohenberg}}]{WHH1966}
\bibinfo{author}{\bibfnamefont{N.~R.} \bibnamefont{Werthamer}},
  \bibinfo{author}{\bibfnamefont{E.}~\bibnamefont{Helfand}}, \bibnamefont{and}
  \bibinfo{author}{\bibfnamefont{P.~C.} \bibnamefont{Hohenberg}},
  \bibinfo{journal}{Phys. Rev.} \textbf{\bibinfo{volume}{147}},
  \bibinfo{pages}{295} (\bibinfo{year}{1966}).

\bibitem[{\citenamefont{Matsuda and Shimahara}(2007)}]{matsuda2007FFLO}
\bibinfo{author}{\bibfnamefont{Y.}~\bibnamefont{Matsuda}} \bibnamefont{and}
  \bibinfo{author}{\bibfnamefont{H.}~\bibnamefont{Shimahara}},
  \bibinfo{journal}{Journal of the Physical Society of Japan}
  \textbf{\bibinfo{volume}{76}}, \bibinfo{pages}{051005}
  (\bibinfo{year}{2007}).

\bibitem[{\citenamefont{Fang et~al.}(2010)\citenamefont{Fang, Yang, Balakirev,
  Kohama, Singleton, Qian, Mao, Wang, and Yuan}}]{Fang2010Weak}
\bibinfo{author}{\bibfnamefont{M.}~\bibnamefont{Fang}},
  \bibinfo{author}{\bibfnamefont{J.}~\bibnamefont{Yang}},
  \bibinfo{author}{\bibfnamefont{F.~F.} \bibnamefont{Balakirev}},
  \bibinfo{author}{\bibfnamefont{Y.}~\bibnamefont{Kohama}},
  \bibinfo{author}{\bibfnamefont{J.}~\bibnamefont{Singleton}},
  \bibinfo{author}{\bibfnamefont{B.}~\bibnamefont{Qian}},
  \bibinfo{author}{\bibfnamefont{Z.~Q.} \bibnamefont{Mao}},
  \bibinfo{author}{\bibfnamefont{H.}~\bibnamefont{Wang}}, \bibnamefont{and}
  \bibinfo{author}{\bibfnamefont{H.~Q.} \bibnamefont{Yuan}},
  \bibinfo{journal}{Phys. Rev. B} \textbf{\bibinfo{volume}{81}},
  \bibinfo{pages}{020509} (\bibinfo{year}{2010}).

\bibitem[{\citenamefont{Khim et~al.}(2010)\citenamefont{Khim, Kim, Choi, Bang,
  Nohara, Takagi, and Kim}}]{khim2010evidence}
\bibinfo{author}{\bibfnamefont{S.}~\bibnamefont{Khim}},
  \bibinfo{author}{\bibfnamefont{J.~W.} \bibnamefont{Kim}},
  \bibinfo{author}{\bibfnamefont{E.~S.} \bibnamefont{Choi}},
  \bibinfo{author}{\bibfnamefont{Y.}~\bibnamefont{Bang}},
  \bibinfo{author}{\bibfnamefont{M.}~\bibnamefont{Nohara}},
  \bibinfo{author}{\bibfnamefont{H.}~\bibnamefont{Takagi}}, \bibnamefont{and}
  \bibinfo{author}{\bibfnamefont{K.~H.} \bibnamefont{Kim}},
  \bibinfo{journal}{Physical Review B} \textbf{\bibinfo{volume}{81}},
  \bibinfo{pages}{184511} (\bibinfo{year}{2010}).

\bibitem[{\citenamefont{Her et~al.}(2015)\citenamefont{Her, Kohama, Matsuda,
  Kindo, Yang, Chareev, Mitrofanova, Volkova, Vasiliev, and
  Lin}}]{her2015anisotropy}
\bibinfo{author}{\bibfnamefont{J.}~\bibnamefont{Her}},
  \bibinfo{author}{\bibfnamefont{Y.}~\bibnamefont{Kohama}},
  \bibinfo{author}{\bibfnamefont{Y.}~\bibnamefont{Matsuda}},
  \bibinfo{author}{\bibfnamefont{K.}~\bibnamefont{Kindo}},
  \bibinfo{author}{\bibfnamefont{W.}~\bibnamefont{Yang}},
  \bibinfo{author}{\bibfnamefont{D.}~\bibnamefont{Chareev}},
  \bibinfo{author}{\bibfnamefont{E.}~\bibnamefont{Mitrofanova}},
  \bibinfo{author}{\bibfnamefont{O.}~\bibnamefont{Volkova}},
  \bibinfo{author}{\bibfnamefont{A.}~\bibnamefont{Vasiliev}}, \bibnamefont{and}
  \bibinfo{author}{\bibfnamefont{J.-Y.} \bibnamefont{Lin}},
  \bibinfo{journal}{Superconductor Science and Technology}
  \textbf{\bibinfo{volume}{28}}, \bibinfo{pages}{045013}
  (\bibinfo{year}{2015}).

\bibitem[{\citenamefont{Nakamura et~al.}(2019)\citenamefont{Nakamura, Adachi,
  Omori, Koike, and Takeyama}}]{nakamura2019pauli}
\bibinfo{author}{\bibfnamefont{D.}~\bibnamefont{Nakamura}},
  \bibinfo{author}{\bibfnamefont{T.}~\bibnamefont{Adachi}},
  \bibinfo{author}{\bibfnamefont{K.}~\bibnamefont{Omori}},
  \bibinfo{author}{\bibfnamefont{Y.}~\bibnamefont{Koike}}, \bibnamefont{and}
  \bibinfo{author}{\bibfnamefont{S.}~\bibnamefont{Takeyama}},
  \bibinfo{journal}{Scientific Reports} \textbf{\bibinfo{volume}{9}},
  \bibinfo{pages}{16949} (\bibinfo{year}{2019}).

\bibitem[{\citenamefont{Wang et~al.}(2021)\citenamefont{Wang, Li, Goodge, Lee,
  Osada, Harvey, Kourkoutis, Beasley, and Hwang}}]{wang2021isotropic}
\bibinfo{author}{\bibfnamefont{B.~Y.} \bibnamefont{Wang}},
  \bibinfo{author}{\bibfnamefont{D.}~\bibnamefont{Li}},
  \bibinfo{author}{\bibfnamefont{B.~H.} \bibnamefont{Goodge}},
  \bibinfo{author}{\bibfnamefont{K.}~\bibnamefont{Lee}},
  \bibinfo{author}{\bibfnamefont{M.}~\bibnamefont{Osada}},
  \bibinfo{author}{\bibfnamefont{S.~P.} \bibnamefont{Harvey}},
  \bibinfo{author}{\bibfnamefont{L.~F.} \bibnamefont{Kourkoutis}},
  \bibinfo{author}{\bibfnamefont{M.~R.} \bibnamefont{Beasley}},
  \bibnamefont{and} \bibinfo{author}{\bibfnamefont{H.~Y.} \bibnamefont{Hwang}},
  \bibinfo{journal}{Nature Physics} \textbf{\bibinfo{volume}{17}},
  \bibinfo{pages}{473} (\bibinfo{year}{2021}).

\bibitem[{\citenamefont{Agosta et~al.}(2017)\citenamefont{Agosta, Fortune,
  Hannahs, Gu, Liang, Park, and Schleuter}}]{Agosta2017organicFFLO}
\bibinfo{author}{\bibfnamefont{C.~C.} \bibnamefont{Agosta}},
  \bibinfo{author}{\bibfnamefont{N.~A.} \bibnamefont{Fortune}},
  \bibinfo{author}{\bibfnamefont{S.~T.} \bibnamefont{Hannahs}},
  \bibinfo{author}{\bibfnamefont{S.}~\bibnamefont{Gu}},
  \bibinfo{author}{\bibfnamefont{L.}~\bibnamefont{Liang}},
  \bibinfo{author}{\bibfnamefont{J.-H.} \bibnamefont{Park}}, \bibnamefont{and}
  \bibinfo{author}{\bibfnamefont{J.~A.} \bibnamefont{Schleuter}},
  \bibinfo{journal}{Phys. Rev. Lett.} \textbf{\bibinfo{volume}{118}},
  \bibinfo{pages}{267001} (\bibinfo{year}{2017}).

\bibitem[{\citenamefont{Cao et~al.}(2018)\citenamefont{Cao, Fatemi, Fang,
  Watanabe, Taniguchi, Kaxiras, and Jarillo-Herrero}}]{cao2018unconventional}
\bibinfo{author}{\bibfnamefont{Y.}~\bibnamefont{Cao}},
  \bibinfo{author}{\bibfnamefont{V.}~\bibnamefont{Fatemi}},
  \bibinfo{author}{\bibfnamefont{S.}~\bibnamefont{Fang}},
  \bibinfo{author}{\bibfnamefont{K.}~\bibnamefont{Watanabe}},
  \bibinfo{author}{\bibfnamefont{T.}~\bibnamefont{Taniguchi}},
  \bibinfo{author}{\bibfnamefont{E.}~\bibnamefont{Kaxiras}}, \bibnamefont{and}
  \bibinfo{author}{\bibfnamefont{P.}~\bibnamefont{Jarillo-Herrero}},
  \bibinfo{journal}{Nature} \textbf{\bibinfo{volume}{556}}, \bibinfo{pages}{43}
  (\bibinfo{year}{2018}).

\bibitem[{\citenamefont{Wan et~al.}(2023)\citenamefont{Wan, Zheliuk, Yuan,
  Peng, Zhang, Liang, Zeitler, Wiedmann, Hussey, Palstra
  et~al.}}]{wan2023orbital}
\bibinfo{author}{\bibfnamefont{P.}~\bibnamefont{Wan}},
  \bibinfo{author}{\bibfnamefont{O.}~\bibnamefont{Zheliuk}},
  \bibinfo{author}{\bibfnamefont{N.~F.} \bibnamefont{Yuan}},
  \bibinfo{author}{\bibfnamefont{X.}~\bibnamefont{Peng}},
  \bibinfo{author}{\bibfnamefont{L.}~\bibnamefont{Zhang}},
  \bibinfo{author}{\bibfnamefont{M.}~\bibnamefont{Liang}},
  \bibinfo{author}{\bibfnamefont{U.}~\bibnamefont{Zeitler}},
  \bibinfo{author}{\bibfnamefont{S.}~\bibnamefont{Wiedmann}},
  \bibinfo{author}{\bibfnamefont{N.~E.} \bibnamefont{Hussey}},
  \bibinfo{author}{\bibfnamefont{T.~T.} \bibnamefont{Palstra}},
  \bibnamefont{et~al.}, \bibinfo{journal}{Nature}
  \textbf{\bibinfo{volume}{619}}, \bibinfo{pages}{46} (\bibinfo{year}{2023}).

\bibitem[{\citenamefont{Kasahara et~al.}(2014)\citenamefont{Kasahara,
  Watashige, Hanaguri, Kohsaka, Yamashita, Shimoyama, Mizukami, Endo, Ikeda,
  Aoyama et~al.}}]{kasahara2014field}
\bibinfo{author}{\bibfnamefont{S.}~\bibnamefont{Kasahara}},
  \bibinfo{author}{\bibfnamefont{T.}~\bibnamefont{Watashige}},
  \bibinfo{author}{\bibfnamefont{T.}~\bibnamefont{Hanaguri}},
  \bibinfo{author}{\bibfnamefont{Y.}~\bibnamefont{Kohsaka}},
  \bibinfo{author}{\bibfnamefont{T.}~\bibnamefont{Yamashita}},
  \bibinfo{author}{\bibfnamefont{Y.}~\bibnamefont{Shimoyama}},
  \bibinfo{author}{\bibfnamefont{Y.}~\bibnamefont{Mizukami}},
  \bibinfo{author}{\bibfnamefont{R.}~\bibnamefont{Endo}},
  \bibinfo{author}{\bibfnamefont{H.}~\bibnamefont{Ikeda}},
  \bibinfo{author}{\bibfnamefont{K.}~\bibnamefont{Aoyama}},
  \bibnamefont{et~al.}, \bibinfo{journal}{Proceedings of the National Academy
  of Sciences} \textbf{\bibinfo{volume}{111}}, \bibinfo{pages}{16309}
  (\bibinfo{year}{2014}).

\bibitem[{\citenamefont{Kasahara et~al.}(2020)\citenamefont{Kasahara, Sato,
  Licciardello, {\v{C}}ulo, Arsenijevi{\'c}, Ottenbros, Tominaga, B{\"o}ker,
  Eremin, Shibauchi et~al.}}]{kasahara2020evidence}
\bibinfo{author}{\bibfnamefont{S.}~\bibnamefont{Kasahara}},
  \bibinfo{author}{\bibfnamefont{Y.}~\bibnamefont{Sato}},
  \bibinfo{author}{\bibfnamefont{S.}~\bibnamefont{Licciardello}},
  \bibinfo{author}{\bibfnamefont{M.}~\bibnamefont{{\v{C}}ulo}},
  \bibinfo{author}{\bibfnamefont{S.}~\bibnamefont{Arsenijevi{\'c}}},
  \bibinfo{author}{\bibfnamefont{T.}~\bibnamefont{Ottenbros}},
  \bibinfo{author}{\bibfnamefont{T.}~\bibnamefont{Tominaga}},
  \bibinfo{author}{\bibfnamefont{J.}~\bibnamefont{B{\"o}ker}},
  \bibinfo{author}{\bibfnamefont{I.}~\bibnamefont{Eremin}},
  \bibinfo{author}{\bibfnamefont{T.}~\bibnamefont{Shibauchi}},
  \bibnamefont{et~al.}, \bibinfo{journal}{Physical Review Letters}
  \textbf{\bibinfo{volume}{124}}, \bibinfo{pages}{107001}
  (\bibinfo{year}{2020}).

\bibitem[{\citenamefont{Jin et~al.}(2011)\citenamefont{Jin, Butch, Kirshenbaum,
  Paglione, and Greene}}]{jin2011link}
\bibinfo{author}{\bibfnamefont{K.}~\bibnamefont{Jin}},
  \bibinfo{author}{\bibfnamefont{N.}~\bibnamefont{Butch}},
  \bibinfo{author}{\bibfnamefont{K.}~\bibnamefont{Kirshenbaum}},
  \bibinfo{author}{\bibfnamefont{J.}~\bibnamefont{Paglione}}, \bibnamefont{and}
  \bibinfo{author}{\bibfnamefont{R.}~\bibnamefont{Greene}},
  \bibinfo{journal}{Nature} \textbf{\bibinfo{volume}{476}}, \bibinfo{pages}{73}
  (\bibinfo{year}{2011}).

\bibitem[{\citenamefont{Tomita et~al.}(2015)\citenamefont{Tomita, Kuga,
  Uwatoko, Coleman, and Nakatsuji}}]{tomita2015strange}
\bibinfo{author}{\bibfnamefont{T.}~\bibnamefont{Tomita}},
  \bibinfo{author}{\bibfnamefont{K.}~\bibnamefont{Kuga}},
  \bibinfo{author}{\bibfnamefont{Y.}~\bibnamefont{Uwatoko}},
  \bibinfo{author}{\bibfnamefont{P.}~\bibnamefont{Coleman}}, \bibnamefont{and}
  \bibinfo{author}{\bibfnamefont{S.}~\bibnamefont{Nakatsuji}},
  \bibinfo{journal}{Science} \textbf{\bibinfo{volume}{349}},
  \bibinfo{pages}{506} (\bibinfo{year}{2015}).

\bibitem[{\citenamefont{Cao et~al.}(2020)\citenamefont{Cao, Chowdhury,
  Rodan-Legrain, Rubies-Bigorda, Watanabe, Taniguchi, Senthil, and
  Jarillo-Herrero}}]{cao2020strange}
\bibinfo{author}{\bibfnamefont{Y.}~\bibnamefont{Cao}},
  \bibinfo{author}{\bibfnamefont{D.}~\bibnamefont{Chowdhury}},
  \bibinfo{author}{\bibfnamefont{D.}~\bibnamefont{Rodan-Legrain}},
  \bibinfo{author}{\bibfnamefont{O.}~\bibnamefont{Rubies-Bigorda}},
  \bibinfo{author}{\bibfnamefont{K.}~\bibnamefont{Watanabe}},
  \bibinfo{author}{\bibfnamefont{T.}~\bibnamefont{Taniguchi}},
  \bibinfo{author}{\bibfnamefont{T.}~\bibnamefont{Senthil}}, \bibnamefont{and}
  \bibinfo{author}{\bibfnamefont{P.}~\bibnamefont{Jarillo-Herrero}},
  \bibinfo{journal}{Physical Review Letters} \textbf{\bibinfo{volume}{124}},
  \bibinfo{pages}{076801} (\bibinfo{year}{2020}).

\bibitem[{\citenamefont{Zhao et~al.}(2019)\citenamefont{Zhao, Zhang, Lyu,
  Bachus, Tokiwa, Gegenwart, Zhang, Cheng, Yang, Chen
  et~al.}}]{zhao2019quantum}
\bibinfo{author}{\bibfnamefont{H.}~\bibnamefont{Zhao}},
  \bibinfo{author}{\bibfnamefont{J.}~\bibnamefont{Zhang}},
  \bibinfo{author}{\bibfnamefont{M.}~\bibnamefont{Lyu}},
  \bibinfo{author}{\bibfnamefont{S.}~\bibnamefont{Bachus}},
  \bibinfo{author}{\bibfnamefont{Y.}~\bibnamefont{Tokiwa}},
  \bibinfo{author}{\bibfnamefont{P.}~\bibnamefont{Gegenwart}},
  \bibinfo{author}{\bibfnamefont{S.}~\bibnamefont{Zhang}},
  \bibinfo{author}{\bibfnamefont{J.}~\bibnamefont{Cheng}},
  \bibinfo{author}{\bibfnamefont{Y.-f.} \bibnamefont{Yang}},
  \bibinfo{author}{\bibfnamefont{G.}~\bibnamefont{Chen}}, \bibnamefont{et~al.},
  \bibinfo{journal}{Nature Physics} \textbf{\bibinfo{volume}{15}},
  \bibinfo{pages}{1261} (\bibinfo{year}{2019}).

\bibitem[{\citenamefont{Wakamatsu et~al.}(2023)\citenamefont{Wakamatsu, Suzuki,
  Fujii, Miyagawa, Taniguchi, and Kanoda}}]{wakamatsu2023thermoelectric}
\bibinfo{author}{\bibfnamefont{K.}~\bibnamefont{Wakamatsu}},
  \bibinfo{author}{\bibfnamefont{Y.}~\bibnamefont{Suzuki}},
  \bibinfo{author}{\bibfnamefont{T.}~\bibnamefont{Fujii}},
  \bibinfo{author}{\bibfnamefont{K.}~\bibnamefont{Miyagawa}},
  \bibinfo{author}{\bibfnamefont{H.}~\bibnamefont{Taniguchi}},
  \bibnamefont{and} \bibinfo{author}{\bibfnamefont{K.}~\bibnamefont{Kanoda}},
  \bibinfo{journal}{Nature Communications} \textbf{\bibinfo{volume}{14}},
  \bibinfo{pages}{3679} (\bibinfo{year}{2023}).

\bibitem[{\citenamefont{Momma and Izumi}(2011)}]{VESTA}
\bibinfo{author}{\bibfnamefont{K.}~\bibnamefont{Momma}} \bibnamefont{and}
  \bibinfo{author}{\bibfnamefont{F.}~\bibnamefont{Izumi}}, \bibinfo{journal}{J.
  Appl. Crystallogr.} \textbf{\bibinfo{volume}{44}}, \bibinfo{pages}{1272}
  (\bibinfo{year}{2011}).

\end{thebibliography}

\begin{center}
\begin{figure*}[tb]
\includegraphics[scale=0.88]{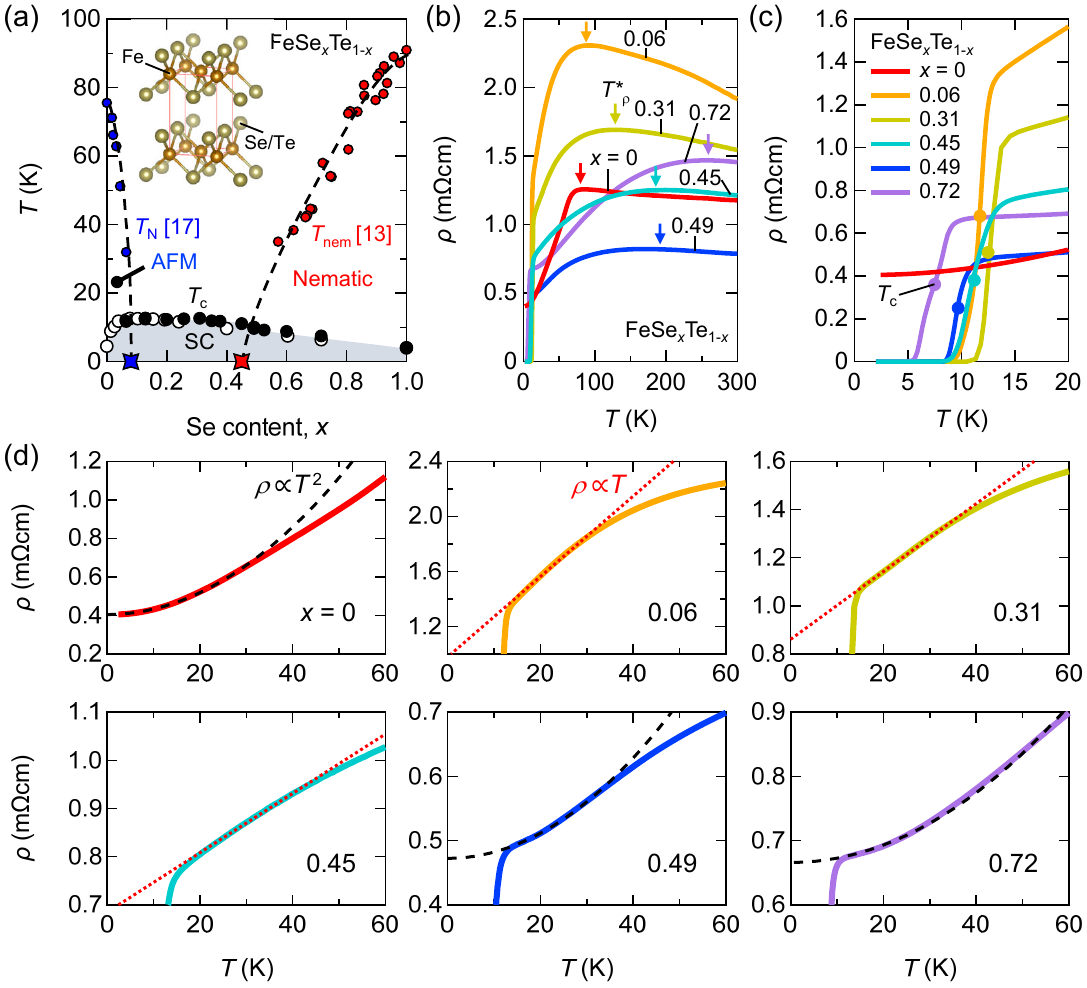}
\caption{\label{fig:RT} \textbf{Non-Fermi liquid (NFL) transport signature in FST thin films.} (a) $T$-$x$ phase diagram of \FST\ (FST). The black and white markers represent \tc\ for FST thin films grown on STO and CdTe substrates \cite{Sato2024Molecular}, respectively. The red and blue markers respectively denote $T_{\rm nem}$ \cite{mukasa2021high} and $T_{\rm N}$ \cite{otsuka2019incoherent}, adopted from resistivity measurements on bulk single crystals. The inset shows the crystal structure of FST in the tetragonal phase \cite{VESTA}. (b) $\rho$-$T$ curves for FST thin films with various $x$ ranging from 0 to 0.72. Arrows indicate $T^{*}_{\rho}$, where $\rho$ reaches its maximum. (c) Low-temperature expansion of (b). Markers denote $T_{\rm c}$ for each sample. (d) Low-temperature normal state resistivity below 60 K. Red dashed lines show linear fits ($\rho_0 + A_1T$), while black dashed lines represent quadratic fits ($\rho_0 + A_2T^2$) in the low-temperature region above \tc.
}
\end{figure*}

\begin{figure*}[tb]
\includegraphics[scale=0.7]{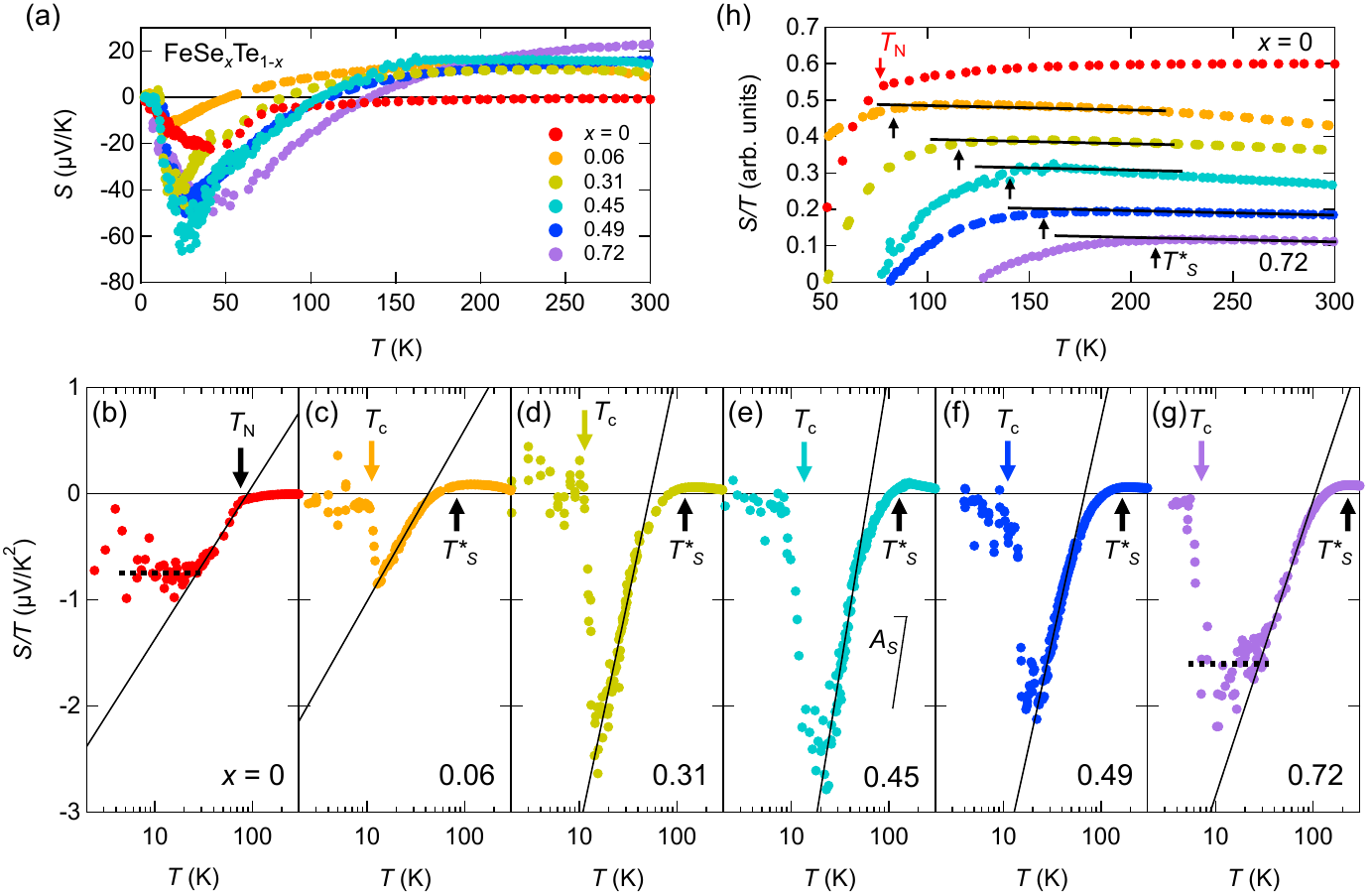}
\caption{\label{fig:ST} \textbf{Thermoelectric signatures fo NFL transport behavior in FST thin films.} (a) $T$-dependence of $S$ for FST thin films with $x$ = 0-0.72. (b)-(g) $S/T$ plotted against logarithmic scale of $T$ for various $x$. The solid lines represent linear fits indicating logarithmic divergence of $S/T$, with $A_S$ denoting the slope of the fits. For $x$ = 0 and 0.72, dashed lines show the saturation of $S/T$. $T_{\rm c}$ and $T_{\rm N}$, determined from $\rho$-$T$ curves, are also shown. (h) High-temperature expansion of $S/T$ with offsets for clarity. At high temperatures, $S/T$ shows weak dependence on $T$, as indicated by the black lines. $T_{S}^{*}$ is defined as the temperature below which $S/T$ deviates from these lines for 0.06 $\leq x \leq$ 0.72. For $x$ = 0, $S/T$ exhibits a kink anomaly at $T_{\rm N}$.}
\end{figure*}

\begin{figure*}[tb]
\includegraphics[scale=0.8]{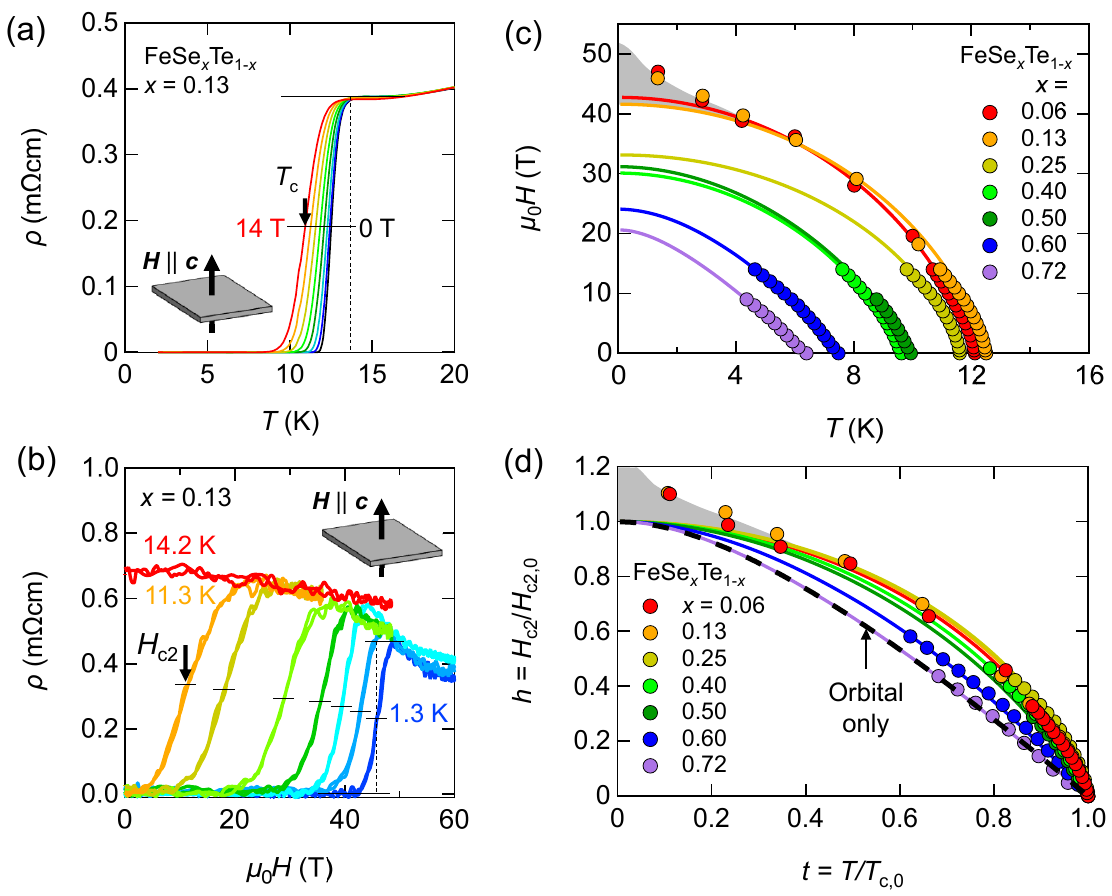}
\caption{\label{fig:Hc2} \textbf{Crossover of pair-breaking mechanisms from orbital to Pauli effect.} (a) $\rho$-$T$ curves for an FST ($x$ = 0.13) thin film under out-of plane magnetic fields, shown in 2 T intervals. $T_{\rm c}$ is defined as the temperature where $\rho$ becomes half of that in the normal state (50\% criterion). (b) $\rho$-$H$ curves at $T$ = 1.3, 2.9, 4.2, 6.0, 8.1, 10.2, 11.3, and 14.2 K. $H_{\rm c2}$ is determined using the 50\% criterion. (c) $H$-$T$ phase diagram for FST with 0.06 $\leq x \leq$ 0.72. The solid curves are fits to the WHH formula incorporating the orbital, Pauli, and spin-orbit scattering terms, with $\lambda_{\rm SO}$ fixed at 1.0. The grey area indicates the region where $H_{\rm c2}$ deviates from the fits, suggesting a possible high-field superconducting phase. (d) Same plot as (c) but plotted against the reduced parameters \textbf{h} and \textbf{t}. The dashed line denotes the WHH formula, incorporating only the orbital effect.}
\end{figure*}

\begin{figure*}[tb]
\includegraphics[scale=0.9]{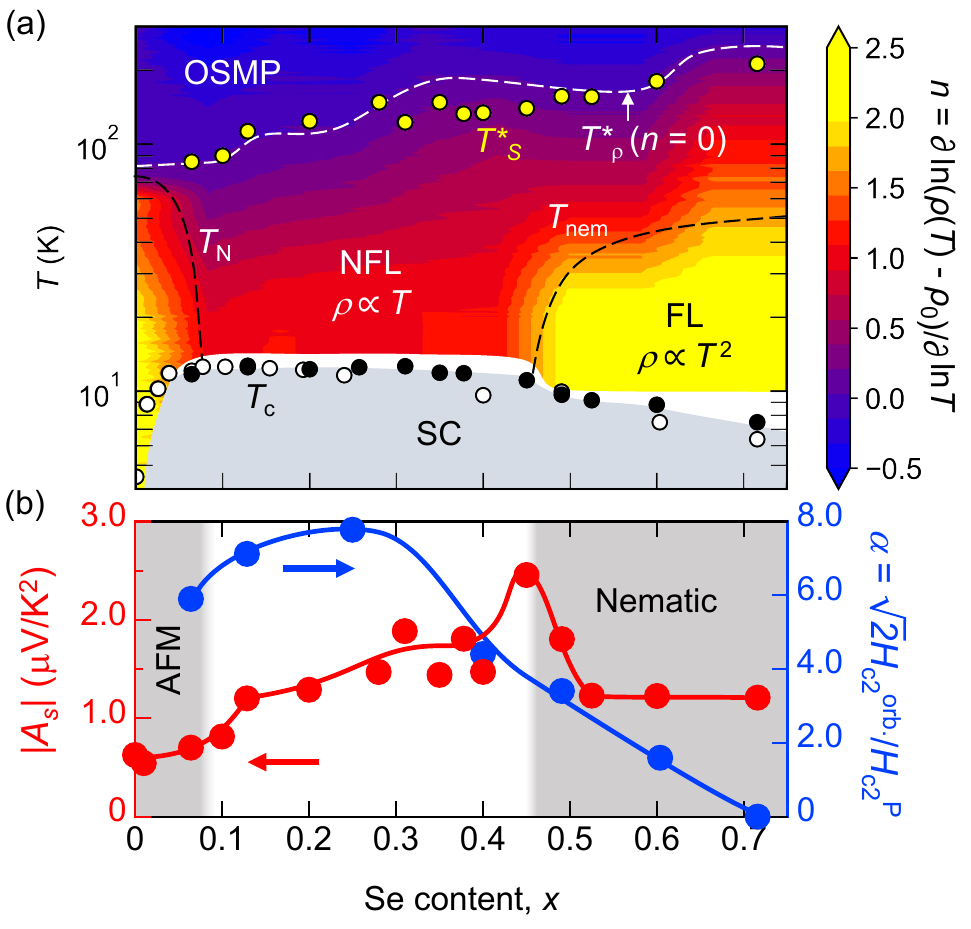}
\caption{\label{fig:phase} \textbf{NFL transport and mass enhancement over the phase diagram.} (a) $T$-$x$ phase diagram with a color plot showing the resistivity power $n (= \partial\ln{(\rho(T) - \rho_0)/\partial\ln{T}})$. The yellow, red, and blue regions represent Fermi liquid (FL) ($n \sim 2$), non-Fermi liquid (NFL) ($0 < n < 2$), and orbital-selective Mott phase (OSMP) ($n \leq 0$), respectively. The yellow markers denote $T_{S}^{*}$ estimated from the Seebeck coefficient measurements. The white dashed line represents the $n = 0$ contour where $\rho$ takes its maximum. The black and white markers indicate $T_{\rm c}$ for FST thin films grown on CdTe and STO substrates, respectively. The black dashed lines represent $T_{\rm N}$ and $T_{\rm nem}$. (b) $x$ dependence of the logarithmic growth of thermoelectricity, $|A_S|$ (left axis), and the Maki parameter $\alpha$ (right axis). The solid lines are guides for the eyes.}
\end{figure*}
\end{center}

\end{document}


\title{Supplemental Material: Non-Fermi liquid transport and \\
strong mass enhancement near the nematic quantum critical point\\
in FeSe$_x$Te$_{1-x}$ thin films}

\author{Yuki Sato$^{1}$}
\email{yuki.sato.yj@riken.jp}
\author{Soma Nagahama$^{2}$}
\author{Ilya Belopolski$^{1}$}
\author{Ryutaro Yoshimi$^{1,3}$}
\author{Minoru Kawamura$^{1}$}
\author{Atsushi Tsukazaki$^{2,4}$}
\author{Akiyoshi Yamada$^{1,5}$}
\author{Masashi Tokunaga$^{1,5}$}
\author{Naoya Kanazawa$^{6}$}
\author{Kei S. Takahashi$^{1}$}
\author{Yoshichika $\bar{\rm O}$nuki$^{1}$}
\author{Masashi Kawasaki$^{1, 2}$}
\author{Yoshinori Tokura$^{1, 2, 7}$}

\affiliation{$^{1}$RIKEN Center for Emergent Matter Science (CEMS), Wako 351-0198, Japan} 
\affiliation{$^{2}$Department of Applied Physics and Quantum-phase Electronics Center (QPEC), University of Tokyo, Tokyo 113-8656, Japan}
\affiliation{$^{3}$Department of Advanced Materials Science, University of Tokyo, Kashiwa 277-8561, Japan}
\affiliation{$^{4}$Institute for Materials Research (IMR), Tohoku University, Sendai 980-8577, Japan}
\affiliation{$^{5}$Institute for Solid State Physics, University of Tokyo, Kashiwa 277-8581, Japan}
\affiliation{$^{6}$Institute of Industrial Science, University of Tokyo, Tokyo 153-8505, Japan}
\affiliation{$^{7}$Tokyo College, University of Tokyo, Tokyo 113-8656, Japan}

\date{\today}
\maketitle

\section{Sample fabrication}
All the FST thin films used in this study were synthesised using the MBE method \cite{Sato2024Molecular}. The thickness of all samples was fixed at 40 nm. The films were synthesised at a substrate temperature of 240$^{\circ}$C in an ultrahigh vacuum chamber, with simultaneous beam fluxes of Fe, Te, and Se. The Se/Te ratio was controlled by adjusting beam fluxes, and the $x$ value was calibrated using inductively coupled plasma mass spectrometry (ICP-MS) analysis. After fabrication, we performed annealing under Te flux to optimize \tc\ and form in-situ FeTe$_2$ capping layer \cite{Sato2024Molecular}. Quantitative estimation of excess iron for the annealed samples via ICP-MS is challenging due to the presence of the capping layer. However, the transport properties of Te-annealed FST thin films are similar to those of single crystals, suggesting a minimal amount of excess iron in the films. The lattice constant $c$ was characterised by x-ray diffraction (Fig. S\ref{fig:Vegard}). $c$ of the MBE-grown thin films exhibits nearly linear dependence on $x$, which is consistent with the behaviour observed in bulk single crystals \cite{mukasa2021high, sun2016influence}. Samples for electrical resistivity measurements were grown on SrTiO$_3$ (STO)(100) substrates, and the identical samples were used for the thermoelectric experiments. Samples for upper critical field measurements were grown on CdTe(100) substrates. FST films grown on the two different substrates show similar \tc-$x$ dependence, with an optimal \tc\ around $x$ = 0.1-0.3 (Fig. 1a). We also observed anomalies in $\rho$-$T$ curves of FeSe and FeTe, showing a good agreement with the nematic and AFM transitions reported in bulk crystals (Fig. S\ref{fig:nematic_afm}). 

\section{Magnetotransport measurements}
We measured the low-field $\rho$-$T$ curves up to $\mu_0H = $14 T using a Quantum Design Physical Properties Measurement System (PPMS). For high-field $H$-scan data extending up to $\mu_0H \sim$ 60 T, we employed a home-made mid-pulse magnet with a pulse width of approximately 36 ms, located at the Institute for Solid State Physics, University of Tokyo \cite{Mitamura2020}.

\section{Kondo effect}
The electronic structure of FST at high temperatures has been argued to exhibit an orbital-selective Mott phase (OSMP), where Fe 3$d$ orbitals show different characteristics due to the strong Hund coupling: localized 3$d_{xy}$ band and itinerant 3$d_{zx}$ and 3$d_{yz}$ bands centred at $\Gamma$ point \cite{yi2017role}. Recently, a scanning tunneling spectroscopy study revealed Fano-type tunneling spectra and significant mass enhancement in FeTe, suggesting Kondo hybridization between the localized Fe 3$d_{xy}$ band and itinerant Te 5$p_z$ orbitals \cite{kim2023kondo}. Indeed, we found that the insulating behaviour in FST can be substantially described by the Kondo model: $\rho \propto -\log{T}$ (dashed lines in Fig. S\ref{fig:Kondo}). Interestingly, the characteristic temperature $T_{\rho}^{*}$, where $\rho$ shows a broad peak, shifts to lower temperatures with decreasing $x$, which is similar to that observed in bulk single crystals \cite{otsuka2019incoherent}. In a Kondo lattice, the mass renormalisation factor $\gamma$ can be expressed as $\gamma = R\ln{2}/T_{\rm K} \sim 10^4/T_{\rm K}$ [mJ/(K$^2\cdot$mol)], where $R$ is the gas constant and $T_{\rm K}$ is the Kondo temperature, and $T_{\rho}^{*}$ can be interpreted as a coherent temperature below which the conduction electrons acquire a coherent band structure. As $T_{\rho}^{*}$ roughly corresponds to $T_{\rm K}$, the observed trend of suppressed $T_{\rho}^{*}$ with decreasing $x$ suggests that $\gamma$ significantly increases in the Te-rich FST possibly due to the stronger electronic correlations.

\section{MIR limit}
In conventional metals, resistivity saturates in high temperatures when $\ell$ reaches its shortest possible scattering length, determined by the lattice constant (the Mott-Ioffe-Regel (MIR) limit). In contrast, a distinctive feature of bad metals is their unusually large resistivity, which grows beyond the MIR limit, not necessarily but sometimes exhibiting a $T$-linear dependence \cite{phillips2022stranger}. To determine if FST falls into this bad metal regime, we calculate the MIR-limiting resistivity $\rho_{\rm MIR}$ as follows. The Drude formula for a two-dimensional multi-band metal is given by
\begin{equation}
    \sigma = \sum_{N_{\rm FS}}\dfrac{de^2\tau}{m^{*}} \approx \dfrac{N_{\rm FS}}{c}\dfrac{e^2}{h}k_{\rm F}\ell,
\end{equation}
where $N_{\rm FS}$ is the number of Fermi surfaces, $d = k_{\rm F}^2/(2\pi c)$ is two-dimensional carrier density, and $\ell = \tau v_{\rm F} = \hbar k_{\rm F}\tau/m^{*}$. For simplicity, we assume that all the Fermi surfaces are isotropic cylinders with the same $m^{*}$ and $k_{\rm F}$ values. Here, $k_{\rm F}\ell$ presents the metallicity of the system and becomes unity in the MIR regime, leading to 
\begin{equation}
    \rho_{\rm MIR} = \dfrac{c}{N_{\rm FS}}\dfrac{h}{e^2}.
\end{equation}
The upper limit of $\rho_{\rm MIR}$ for FST is obtained by using the maximum $c$ value of 0.63 nm among FST (Fig. S\ref{fig:Vegard}) and the minimum $N_{\rm FS} = 2$ for the Fermi surface topology of iron-based superconductors, yielding $\rho_{\rm MIR} \sim$ 0.8 m$\Omega$cm. We find that the experimental value of $\rho$ is of the same order or even exceeds this limit, demonstrating the intrinsic bad metal behaviour of FST.

\section{Thermoelectricity}
Thermoelectric measurements were conducted using a standard steady-state method with a home-made sample stage and PPMS. One side of the sample was attached to a copper thermal bath, while a 10 k$\Omega$ tip heater was affixed to the opposite side. Applying current to the heater creates thermal gradient $\Delta T$ across the sample, which was sensed using two pairs of type-E thermocouples placed on the hot and cold sides. The longitudinal DC voltage $\Delta V$ generated was measured with a Keithley nanovoltmeter 2182A. The Seebeck coefficient $S$ was calculated as $S = \Delta V/\Delta T$.\\
In the diffusive thermoelectric regime, $S$ of a FL can be described by the Mott expression, given by 
\begin{equation}
    S = \dfrac{\pi^2}{3}\dfrac{1}{e}\dfrac{k_{\rm B}^2T}{E_{\rm F}},
\end{equation}
where $e$ is the elementary charge, $k_{\rm B}$ is the Boltzmann constant. At sufficiently low enough temperature ($T \ll E_{\rm F}/k_{\rm B}$), $S/T$ is expected to be constant in the FL state.
In the vicinity of a QCP, on the other hand, strong scatterings from quantum fluctuations can be renormalized into $m^{*}$, leading to a logarithmic divergence of $S/T$ \cite{Paul2001Thermoelectric}, described by
\begin{equation}
S/T \propto \dfrac{1}{e}\left( \dfrac{g_0^2N'(0)}{E_{\rm F}\omega_sN(0)}\right)\ln{\left(\dfrac{\omega_s}{\delta}\right)},
\end{equation}
where $g_0$ represents the coupling between carriers and spin fluctuations, $\omega_s$ is the energy scale of spin fluctuations, $N(0)$ and $N'(0)$ are the density of states (DOS) and the energy derivative of DOS at Fermi level, respectively, and $\delta$ measures the distance from the QCP in the phase space. For a fixed $x$ in a given sample, $\delta \approx T$, resulting in the $S/T \propto \ln{T}$ dependence.

\section{WHH fittings}
To capture the systematic evolution of \Hco\ and \HcP\ against $x$, we performed analyses based on the WHH formula for a type-II superconductor in the dirty limit, incorporating both orbital and Pauli effects \cite{WHH1966}:
\begin{align}
    \ln{\dfrac{1}{t}} &= \left( \dfrac{1}{2} + \dfrac{i\lambda_{\rm SO}}{4\gamma} \right)\psi\left( \dfrac{1}{2} + \dfrac{\bar{h} + \lambda_{\rm SO}/2 + i\gamma}{2t} \right)\nonumber\\
    &+ \left( \dfrac{1}{2} - \dfrac{i\lambda_{\rm SO}}{4\gamma} \right)\psi\left( \dfrac{1}{2} + \dfrac{\bar{h} + \lambda_{\rm SO}/2 - i\gamma}{2t} \right) - \psi\left( \dfrac{1}{2}\right).
\end{align}
Here, $\psi(x)$ is the digamma function, $ t = T/T_{\rm c,0}$, $\gamma \equiv [(\alpha\bar{h})^{2} - \lambda_{\rm SO}^2]^{1/2}$, and $\lambda_{\rm SO}$ parameterises the strength of spin-orbit scattering. Additionally, 
\begin{equation}
    \bar{h} = \dfrac{4}{\pi^2}\dfrac{H_{\rm c2}}{(-dH_{\rm c2}/dt)_{t=1}}
\end{equation}
is estimated from linear fittings near $t = 1$. For simplicity, we fixed $\lambda_{\rm SO} = 1$, which provides a reasonable fit for $x$ = 0.06 across the entire $x$ range (Fig. S\ref{fig:twostep}). All the data, fitting results, and the fitting parameters are summarized in Figs. S\ref{fig:inplane}-S\ref{fig:Hc2orb}.

Outside of the critical regime, the $H$-$T$ phase diagrams of FST are anisotropic with respect to the field direction, reflecting its quasi-two-dimensional electronic structure (Fig. S\ref{fig:all_HT}). In the nematic critical regime, on the other hand, the $H$-$T$ phase diagrams become nearly isotropic, where the anisotropy in $\xi$ ($\xi_{ab}/\xi_{c}$) derived from the anisotropic Ginzburg-Landau formula approaches unity (Fig. S\ref{fig:xi}), analogous to heavy fermion superconductors.

\nocite{*}
\bibliography{FST_QCP_SM}

\renewcommand{\figurename}{FIG. S}
\newpage
\begin{center}
\begin{figure*}[tb]
\includegraphics[scale=0.7]{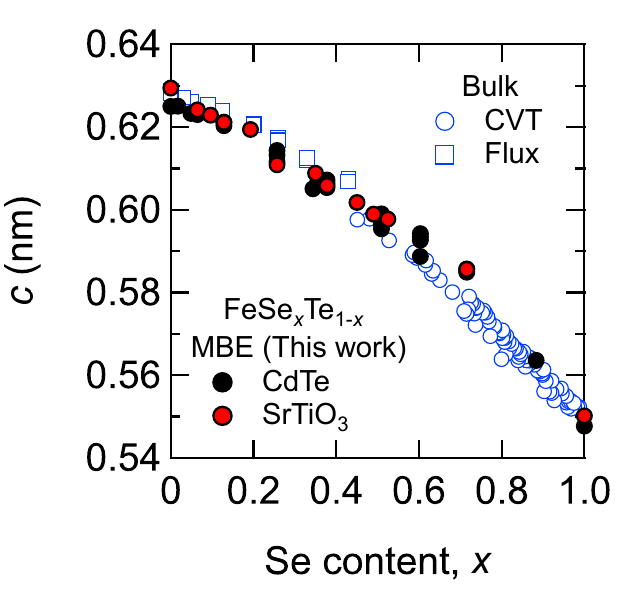}
\caption{\label{fig:Vegard} \textbf{Systematic evolution of lattice constant \textit{c} as a function of \textit{x}.} The lattice constant $c$ is determined by x-ray diffraction measurements at room temperature. The black and red circles represents $c$ for FST thin films grown on CdTe \cite{Sato2024Molecular} and STO substrates, respectively. The blue circles and squares represent $c$ for bulk single crystals grown by CVT \cite{mukasa2021high} and flux techniques \cite{sun2016influence}, respectively.}
\end{figure*}

\begin{figure*}[tb]
\includegraphics[scale=0.7]{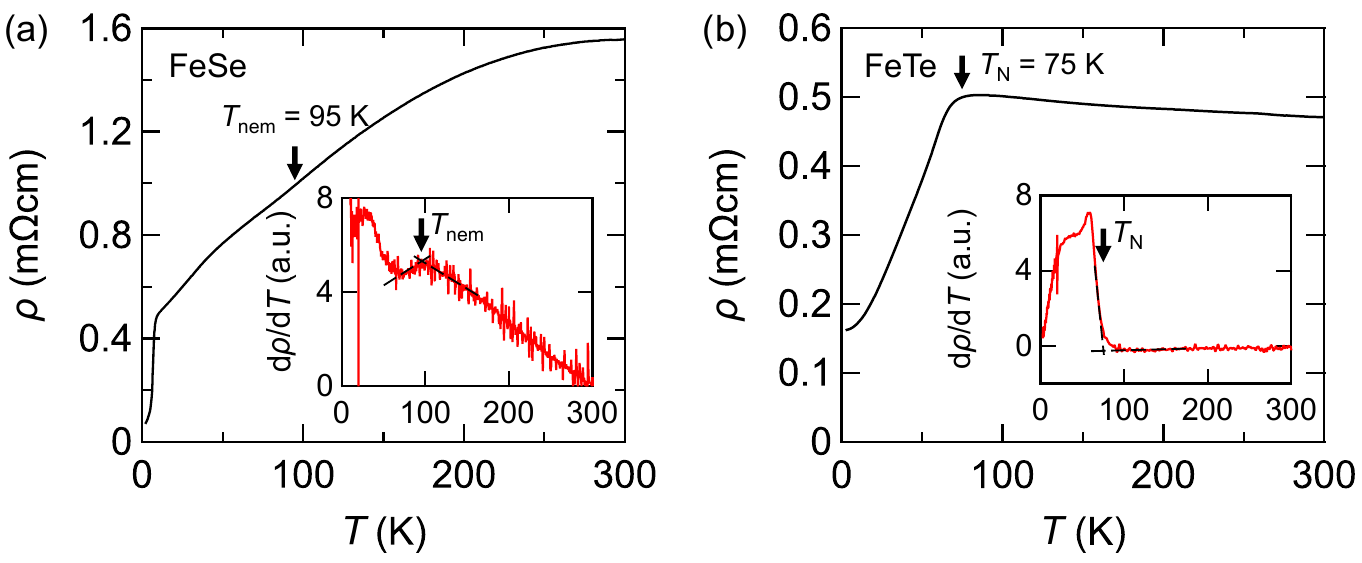}
\caption{\label{fig:nematic_afm} \textbf{Nematic and AFM transitions in the parent compounds FeSe and FeTe thin films.} (a) $T$-dependence of $\rho$ for a FeSe thin films. The inset shows the $T$-derivative of $\rho$. The nematic transition is indicated by a kink in $d\rho/dT$ at approximately $T_{\rm nem} = 95$ K (b) $T$-dependence of $\rho$ for a FeTe thin film. The inset shows $T$-derivative of $\rho$. The AFM transition is marked by a kink in $d\rho/dT$ at around $T_{\rm N} = 75$ K.}
\end{figure*}

\begin{figure*}[tb]
\includegraphics[scale=0.75]{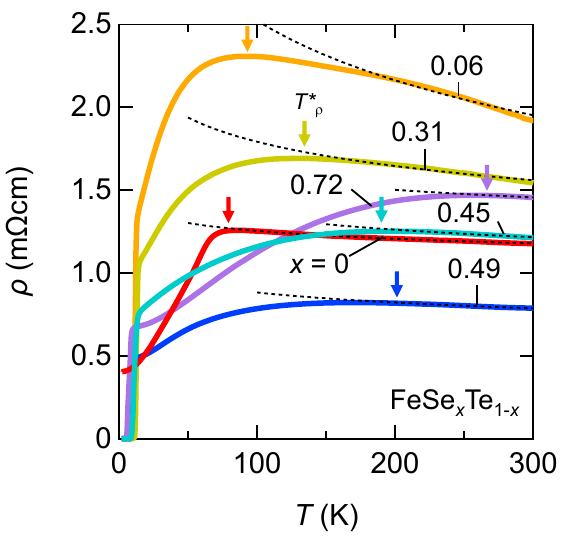}
\caption{\label{fig:Kondo} \textbf{Resistivity fitting with Kondo model.} $\rho$-$T$ curves for FST thin films with various $x$ ranging from 0 to 0.72. The dashed lines represent fits to the Kondo model, $\rho \propto -\log{T}$. Arrows indicate $T_{\rho}^{*}$, the temperature where $\rho$ reaches its maximum.}
\end{figure*}


\begin{figure*}[tb]
\includegraphics[scale=0.6]{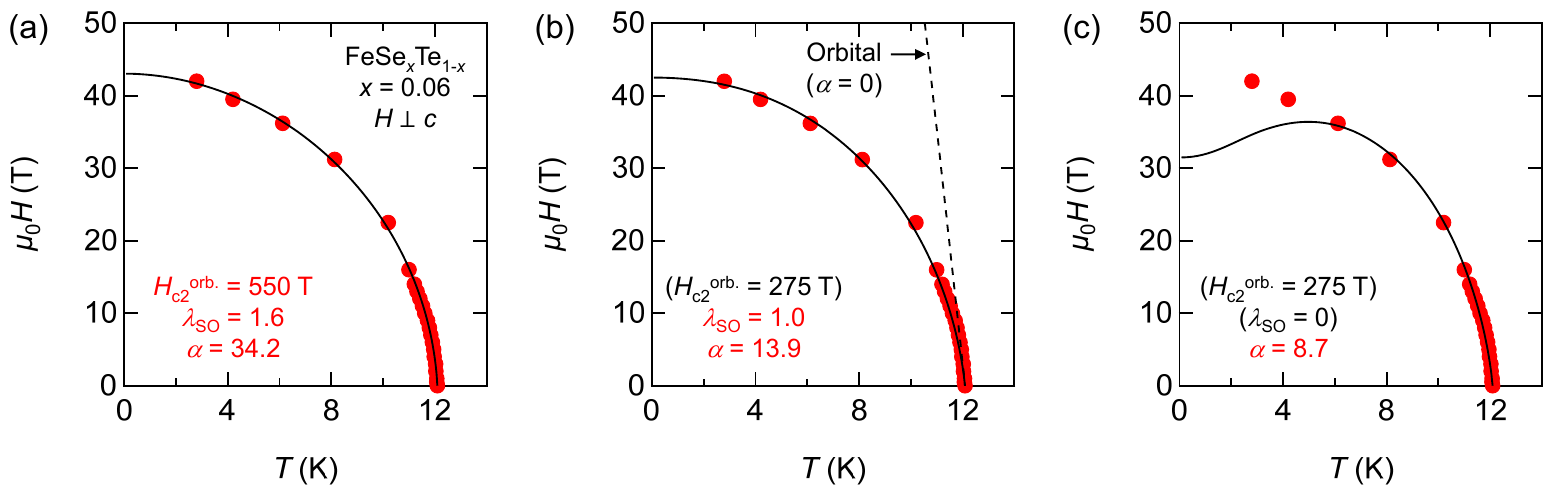}
\caption{\label{fig:twostep} \textbf{Two-step fittings for $H$-$T$ curves with the WHH formula.} (a) Calculation with three fitting parameters: \Hco, $\lambda_{\rm SO}$, and $\alpha$. Representative data with an in-plane field configuration, where the orbital effect is negligible, is used to demonstrate the importance of two-step fittings. This calculation results in a large uncertainty in determining \Hco, as the fitting curve is mainly influenced by paramagnetic effects. (b) The same fitting with \Hco\ fixed at 275 T, obtained from a low-$H$ linear fit below 3 T. The dashed line represents the WHH formula with solely orbital effect. Fitting the low-$H$ data provides a more accurate estimation of \Hco. Here, we obtained $\lambda_{\rm SO} = 1.0$, which is a typical value for all $x$. Thus, in the main text, $\lambda_{\rm SO} = 1.0$ is used for simplicity in the $H$-$T$ curves with the the out-of-plane configuration. (c) Calculation with \Hco\ = fixed at 275 T and $\lambda_{\rm SO}$ fixed at 0. This results in a $H$-$T$ curve with an upward concavity at low $T$. Therefore, inclusion of a finite $\lambda_{\rm SO}$ is necessary for estimation of $\alpha$.}
\end{figure*}


\begin{figure*}[tb]
\includegraphics[scale=0.7]{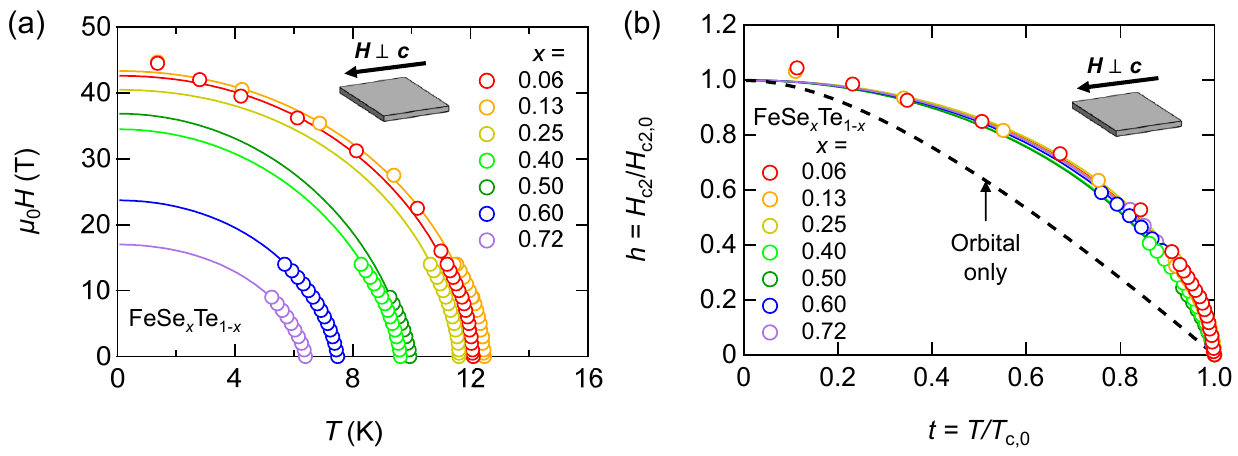}
\caption{\label{fig:inplane} \textbf{$H$-$T$ phase diagram for in-plane field configuration.} (a) $H$-$T$ phase diagram for FST with 0.06 $\leq x \leq$ 0.72 for in-plane field configuration. The solid curves are fits to the WHH formula incorporating the orbital, Pauli, and spin-orbit scattering terms, with $\lambda_{\rm SO}$ fixed at 1.0. (b) Same plot as (b) but plotted against the reduced parameters \textbf{h} and \textbf{t}. The dashed line denotes the WHH formula, incorporating only the orbital effect.}
\end{figure*}

\begin{figure*}[tb]
\includegraphics[scale=0.5]{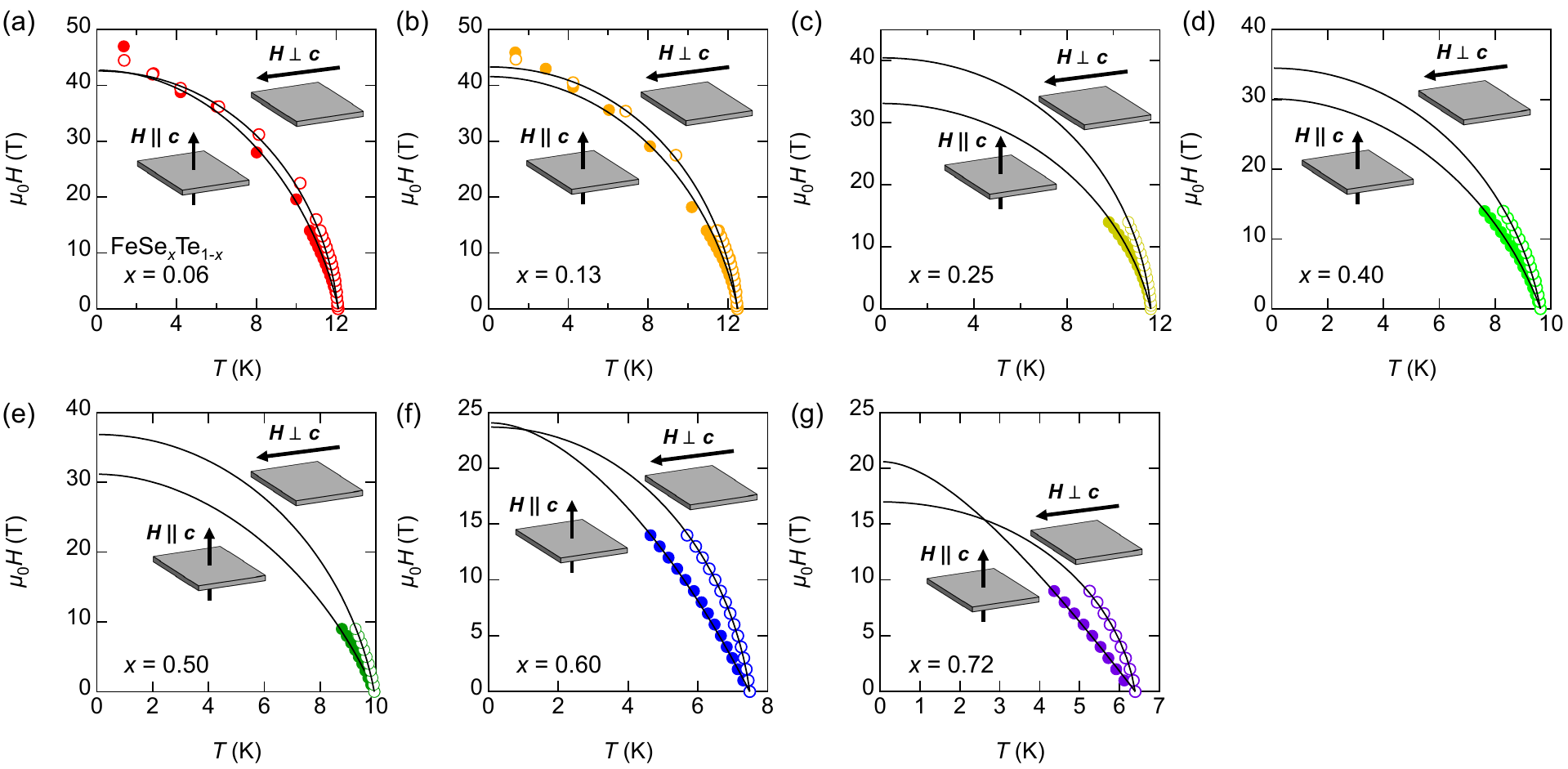}
\caption{\label{fig:all_HT} \textbf{$H$-$T$ phase diagrams of FST with various $x$.} The solid (open) markers represent data taken in out-of-plane (in-plane) field configurations. The solid lines are fits to the WHH formula, incorporating both orbital and Pauli effects.}
\end{figure*}

\begin{figure*}[tb]
\includegraphics[scale=0.53]{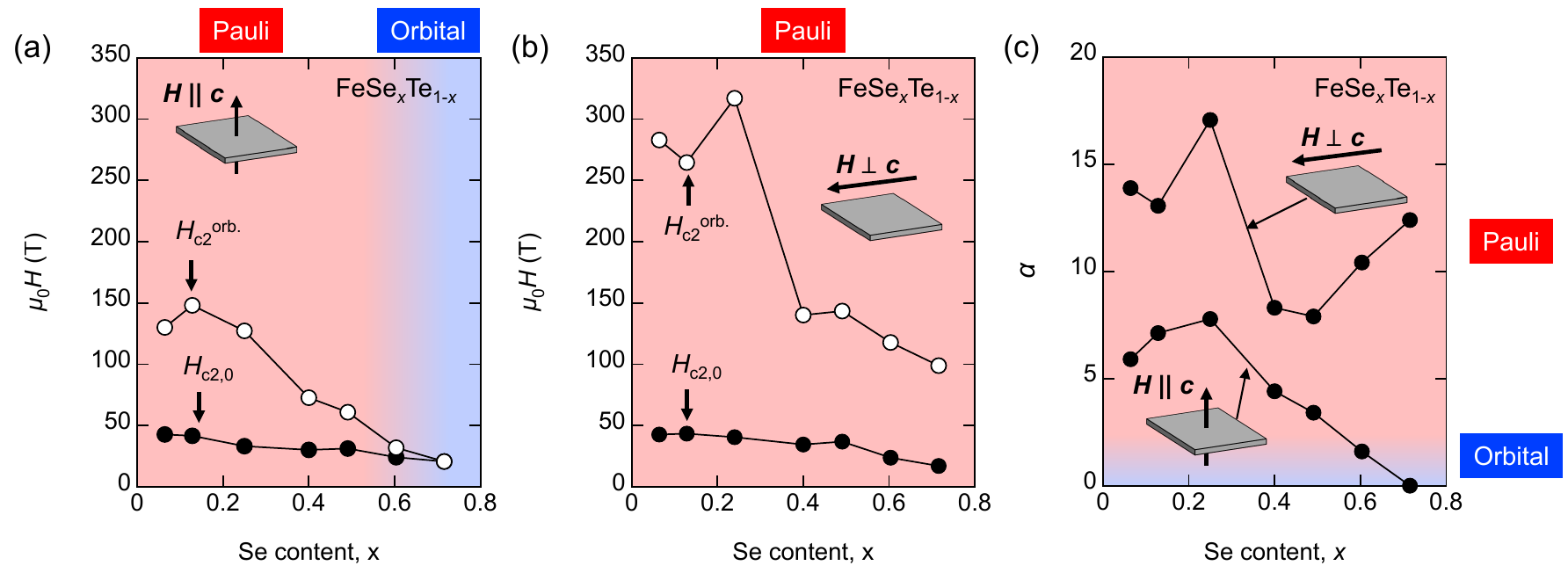}
\caption{\label{fig:Hc2orb} \textbf{Fitting parameters with the WHH formula.} (a) $x$-dependence of \Hco\ and $H_{\rm c20}$ for the out-of-plane configuration. The background colour highlights the predominant pair-breaking mechanisms. In Se-rich compositions ($x \leq 0.6$), \Hco$\approx H_{\rm c20}$, indicating that the orbital effect mainly determines \Hc. In Te-rich compositions ($x \geq 0.3$), \Hco\ is significantly larger than $H_{\rm c20}$, suggesting that Pauli effect plays a substantial role in determining \Hc. (b) $x$-dependence of \Hco\ and $H_{\rm c20}$ for the in-plane configuration. For the entire $x$ range, \Hco\ is consistently above $H_{\rm c20}$, indicating a Pauli-limited \Hc. (c) $x$-dependence of $\alpha$. The Pauli effect dominates ($\alpha > 1.5$) except in the Se-rich samples for the out-of-plane field configuration.}
\end{figure*}

\begin{figure*}[tb]
\includegraphics[scale=0.8]{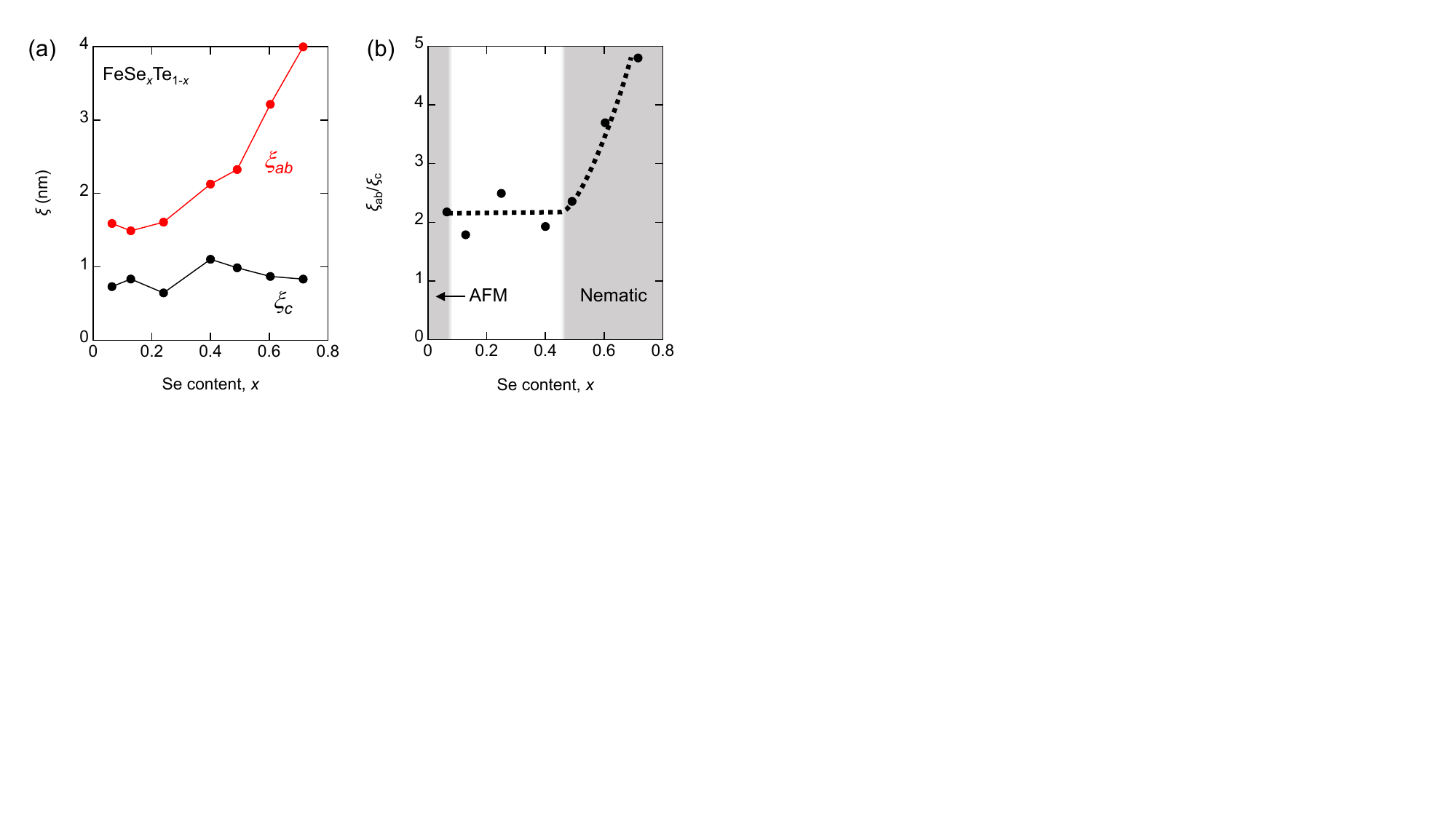}
\caption{\label{fig:xi} \textbf{Coherence length calculated using the anisotropic Ginzburg-Landau formula.} (a) $x$-dependence of coherence length along the in-plane ($\xi_{ab}$) and out-of-plane ($\xi_{c}$) directions. (b) $x$-dependence of the anisotropy ratio $\xi_{ab}$/$\xi_{c}$. The dashed line serves as a guide for the eyes.}
\end{figure*}

\end{center}